\documentclass[useAMS,usenatbib,usegraphicx]{mn2e}
\usepackage{times,amsmath,subfigure}

\title[SDSS Galaxy Colour, Morphology and Environment]{Galaxy Colour,
  Morphology, and Environment in the Sloan Digital Sky Survey}

\author[N. M. Ball et al.]{N. M. Ball,$^{1,2}$\thanks{E-mail:
    nball@astro.uiuc.edu} J. Loveday$^{3}$ and R. J. Brunner$^{1,2}$ \\
  $^{1}$Department of Astronomy, MC-221, University of Illinois, 1002 West Green
  Street, Urbana, IL 61801, USA \\ $^{2}$ National Center for Supercomputing
  Applications, MC-476, University of Illinois, 605 East Springfield Avenue,
  Champaign, IL 61820, USA\\ $^{3}$Astronomy Centre, University of Sussex,
  Falmer, Brighton, BN1 9QJ, UK}

\newcommand{\mrm}{\mathrm}
\newcommand{\omegam}{\Omega_{\mrm{matter}}}
\newcommand{\omegal}{\Omega_{\mrm{\Lambda}}}
\newcommand{\mrlogh}{M_r - 5 {\mrm{log}} h}
\newcommand{\ciinv}{CI_{\mrm{inv}}}
\newcommand{\tann}{T_{\mrm{ANN}}}
\newcommand{\logn}{\mrm{log}~n}
\newcommand{\logsigma}{\mrm{log}~\Sigma_5}
\newcommand{\kms}{~\mrm{km~s^{-1}}}
\newcommand{\kmsp}{~\mrm{km~s^{-1}~Mpc^{-1}}}

\newcommand{\aap}{A\&A}
\newcommand{\aj}{AJ}
\newcommand{\apj}{ApJ}
\newcommand{\apjl}{ApJL}
\newcommand{\apjs}{ApJS}
\newcommand{\araa}{ARA\&A}
\newcommand{\mnras}{MNRAS}
\newcommand{\pasa}{PASA}
\newcommand{\pasp}{PASP}
\newcommand{\physrep}{PhR}

\def\spose#1{\hbox to 0pt{#1\hss}}
\def\simlt{\mathrel{\spose{\lower 3pt\hbox{$\mathchar"218$}}
     \raise 2.0pt\hbox{$\mathchar"13C$}}}
\def\simgt{\mathrel{\spose{\lower 3pt\hbox{$\mathchar"218$}}
     \raise 2.0pt\hbox{$\mathchar"13E$}}}

\begin{document}

\date{Accepted xxxx Received xxxx}

\pagerange{\pageref{firstpage}--\pageref{lastpage}} \pubyear{2007}

\maketitle

\label{firstpage}

\begin{abstract}
We use the Fourth Data Release of the Sloan Digital Sky Survey to investigate
the relation between galaxy rest frame $u-r$ colour, morphology, as described by
the concentration and S\'ersic indices, and environmental density, for a sample
of 79,553 galaxies at $z \simlt 0.1$. We split the samples according to density
and luminosity and recover the expected bimodal distribution in the
colour-morphology plane, shown especially clearly by this subsampling.

We quantify the bimodality by a sum of two Gaussians on the colour and
morphology axes and show that, for the red/early-type population both colour and
morphology do not change significantly as a function of density. For the
blue/late-type population, with increasing density the colour becomes redder but
the morphology again does not change significantly. Both populations become
monotonically redder and of earlier type with increasing luminosity. There is no
significant qualitative difference between the behaviour of the two 
morphological measures.

Motivated by their long-standing use in astronomy and their ability to utilise
information not necessarily used by the concentration and S\'ersic indices, we
supplement the morphological sample with 13,655 galaxies assigned Hubble types
by an artificial neural network. We find, however, that the resulting
distribution is less well described by two Gaussians. Therefore, there
are either more than two significant morphological populations, physical
processes not seen in colour space, or the Hubble type, particularly the
different subtypes of spirals Sa--Sd, has an irreducible fuzziness when related
to environmental density.

For each of the three measures of morphology, on removing the density relation
due to it, we recover a strong residual relation in colour. However, on
similarly removing the colour-density relation there is no evidence for a
residual relation due to morphology. Therefore, either the morphology is not
directly affected by the environmental density beyond the correlation to colour,
or a single galaxy `type' does not capture sufficient information.
\end{abstract}


\section{Introduction} \label{sec: intro} 

The connection between the morphology of a galaxy and the density of its
environment \citep[e.g.][]{oemler:md,dressler:morphdensity}, measured in a
variety of ways, is an important clue to the physics of galaxy formation, and
forms a readily observable quantity that can be compared with simulations. 

Numerous physical processes affect the evolution of a galaxy as a function of
its density environment. The debate is ongoing as to which of these are
intrinsic to the galaxies at formation and which are the result of later
evolution. The relative importance of different processes depends on the density
of the environment. For example, in clusters infalling galaxies are subject to 
ram-pressure stripping and galaxy harassment. In groups, mergers, interactions
and strangulation are more important. It is thought that ellipticals formed
their stars early, exhausting their gas supply and are now passively evolving in
dense environments, whereas spirals formed their stars more slowly, continuing
to the present day and becoming subject to the physical processes described on
infall into denser environments as the structure evolved. We do not attempt to
provide a full reference list for physical processes here, but recent reviews of
the subject include \citet{boselli:latetypecluster}, \citet{hogg:localevoln} and
\citet{avilareese:evolution}.

However, the morphology of the galaxies is affected in ways not necessarily
correlated with the effect on spectral properties or colour. One can therefore
investigate the morphology-density and colour-density relations and subtract the
two to see if a residual relation remains. If so, then there is an intrinsic
effect of density that is not a result of a more fundamental process which
causes the other relation.

A particular example of this is the hypothesis that the properties of galaxies
within a dark matter halo depend only on the mass of the halo
\citep[e.g.][]{cooray:haloreview} This can be tested both for colour and for
morphology, although here we do not specifically relate our results to
haloes. Many results support this assumption
\citep[e.g.][]{blanton:halo,abbas:clustering,blanton:matter,tinker:void}.

Numerous existing results support colour as being a more fundamental predictor
of environment than morphology
\citep[e.g.][]{kauffmann:envt,blanton:envtbbprops,quintero:clustercentric,martinez:envtgroups,quintero:asymmetric}. However,
the question is not yet necessarily settled as \citet{park:envt} find that the
morphology and luminosity are more fundamental. Similarly, \citet{lane:envt}
study the Abell 901/902 supercluster and find that the local environment is more
important than the global for morphology, but \citet{einasto:superclusterenvt}
study numerous superclusters in the 2dF Galaxy Redshift survey
\citep{colless:2df} and find that both the cluster and supercluster environment
influence morphology.

An example of a physical process causing a distinct change in morphology is that
of S0 galaxies. \citet{christlein:fading} find that they must have formed from
bulge enhancement and not disc fading. \citet{boselli:latetypecluster} find that
S0s are formed through gravitational interaction rather than spiral interaction
with the intergalactic medium. \citet{moran:S0s} directly image spirals in the
process of being transformed into S0s.

In the SDSS, the colour-density relation was previously investigated by
\citet[][hereafter B04]{balogh:bimodallfenvt}. They found a pronounced bimodal
distribution well-fit by a sum of two Gaussians. At fixed luminosity, the mean
colours of the two distributions were approximately independent of environment,
but the red fraction strongly increased with density. The density measure used
was the projected distance to the fifth nearest neighbour in a redshift slice of
width $1000 \kms$. The part of our work dealing with $u-r$ significantly
parallels this work, although for a sample size containing three times as many
galaxies.

Previous work in the SDSS on the morphology-density relation includes
\citet{goto:md}, in which they use the concentration index and the texture
parameter of \citet{yamauchi:citexture} to study galaxies in the SDSS Early Data
Release \citep[EDR,][]{stoughton:edr}. \citet{sorrentino:fieldvoid} use the
$u-r$ colour, the eClass eigenclass spectral type and the fraction of the galaxy
profile described by a de Vaucouleurs profile (FracDeV) to describe their
sample, with an emphasis on comparing galaxies in voids to the general
population. \citet{park:envt} use an adaptive kernel estimate of density and the
morphological separation via colour gradient of \citet{park:segregation} and
find little residual dependence of properties on environment at fixed luminosity
and morphology, except for a sharp decrease in late types at the bright
end. Some of these results are described further in \S \ref{subsec: comparison}
below.

Here, we use the Fourth Data Release \citep[DR4,][]{adelmanmccarthy:dr4} of the
Sloan Digital Sky Survey \citep[SDSS,][]{york:sdss} to investigate the
colour-density and morphology-density relations in a large volume-limited
sample. We divide the sample by density and luminosity and investigate the
trends of colour and morphology as a function of each, fitting Gaussians to the
bimodal distributions in colour and morphology.

The sample size, quality of photometry and corresponding spectra enable robust
conclusions to be drawn from the data. However, the available resolution limits
us to simple measures of colour and morphology. Accordingly, as done by previous
authors, we use the $u-r$ colour, inverse concentration index, $\ciinv$ and the
S\'ersic index, $n$, each single-number axisymmetric measures of galaxy
properties. However, in addition to this, we supplement the morphological
measures with the Hubble type assigned by an artificial neural network
\citep[ANN,][]{ball:ann}. This assigns T types 0--6, corresponding to E, S0, Sa,
Sb, Sc, Sd and Im, to a limiting magnitude of $r < 15.9$ (compared to $r <
17.77$ for $u-r$, $\ciinv$ and S\'ersic $n$).

The Hubble type utilises more information from an image, for example spiral arm
structure, and as such may show patterns unseen in the $\ciinv$ and S\'ersic
$n$. Given the wide usage over time of this measure
\citep[e.g.][]{sandage:classfnhistory} it is important that its usefulness is
investigated in this context: if the results using the simple Hubble type used
here are not useful even with the quality and quantity of the data we utilise,
this will emphasize the requirement for different and more detailed measures of
morphology for progress to be made.

This sample used here is the largest to date for which the Hubble type measure
of morphology has been investigated as a function of environment
(c.f. \citet{vandenbergh:envt}, who use the detailed eyeball-assigned Hubble
types for 1,246 galaxies from the Revised Shapley-Ames Catalogue
\citep{sandage:shapleyames}; and \citet{conselice:classification}, who construct
a sample of 22,121 galaxies from the Third Reference Catalog of Bright Galaxies
\citep[RC3,][]{devaucouleurs:rc3}, including Hubble types, but are investigating
ensemble properties rather than environment, which would require a more
restricted sample due in part to the inhomogeneity of the RC3).

Throughout, the standard spatial geometry is assumed, with Euclidean space,
$\Omega_{\mrm{matter}} = 0.3$, $\Omega_{\Lambda} = 0.7$ and dimensionless Hubble
constant $h=1$, where $h = H_0/100 \kmsp$.

\section{Data} \label{sec: data} 

The SDSS is a project to map $\pi$ steradians of the northern galactic cap in
five bands ($u$, $g$, $r$, $i$ and $z$) from 3,500--8,900 \AA. This will provide
photometry for of order $5 \times 10^{7}$ galaxies. A multifibre spectrograph
will provide redshifts and spectra for approximately $10^{6}$ of these. A
technical summary of the survey is given in \citet{york:sdss}. The telescope is
described in \citet{gunn:sdsstelescope}. The imaging camera is described in
\citet{gunn:sdsscamera}. The photometric system and calibration are described in
\citet{fukugita:sdssphotometry}, \citet{hogg:sdssmt}, \citet{smith:ugrizstars},
\citet{ivezic:sdssdata} and \citet{tucker:calibration}. The astrometric
calibration is in \citet{pier:sdssastrometry} and the data pipelines are in
\citet{lupton:sdssphoto}, \citet{lupton:deblender} for the deblender, Frieman et
al. and Schlegel et al. (in preparation).

The targeting pipeline \citep{strauss:mainsample} chooses targets for
spectroscopy from the imaging. A tiling algorithm \citep{blanton:tiling} then
assigns the spectroscopic fibres to the targets, the main source of
incompleteness being the minimum distance of 55 arcsec between the fibres. This
causes about 6\% of galaxies to be missed; those that are will be biased towards
regions with a high surface density of galaxies. The algorithm gives a more
uniform completeness on the sky than a uniform tiling by taking into account
large scale structure, but the effect is still present. The other main source of
incompleteness is galaxies blended with saturated stars, which is a 1\% level
effect. The overall spectroscopic completeness is therefore estimated to be over
90\% \citep{strauss:mainsample}.

The SDSS galaxies with spectra consist of a flux-limited sample, Main, with a
median redshift of 0.104 \citep{strauss:mainsample}; a Luminous Red Galaxy
sample (LRG), approximately volume-limited to $z \approx 0.4$
\citep{eisenstein:lrgsample}; and a quasar sample
\citep{richards:qsosample}. The limiting magnitude for the Main spectra is $r <
17.77$, which is substantially brighter than that for the imaging so the
redshift completeness is almost 100\%. A typical signal-to-noise value is $>4$
per pixel and the spectral resolution is 1800. The redshifts have an RMS
accuracy of $\pm 30 \kms$.

We use galaxies from the Main Galaxy Sample. The data are extracted from the New
York Value-Added Galaxy Catalogue \citep[VAGC,][]{blanton:vagc}. They are
masked, extinction- and K-corrected in the same way as the datasets in
\citet[][hereafter B06]{ball:bivlf}. There is no correction for evolution, but
as the redshifts are limited to $z \simlt 0.1$ its effects are not
large. Similarly, there is no correction for dust, but this is mitigated by the
requirement that the galaxy axis ratio be less than that of an E7 galaxy, as
described in B06.

\subsection{Galaxy Environment} \label{subsec: envt} 

The masking of the dataset as in B06 leaves a sample of 489,123 galaxies. These
are matched to the galaxies in the Pittsburgh-CMU Value Added Catalogue (VAC,
http://\-nvogre.\-phyast.\-pitt.\-edu/\-dr4\_value\_added). The matches were
carried out using the unique identification, both in object and date of
observation, provided by the spectroscopic plate, mjd and fibre values for each
object. The matches were made in order to utilise their measures of the distance
from each galaxy to the nearest survey edge, which use a sophisticated algorithm
taking into account the full survey mask. Galaxies measured by the VAC as being
nearer to a survey edge than their Nth nearest neighbour are excluded to prevent
the densities being biased downward near the survey edge or at the higher
redshifts in the sample.

We do, however, calculate our own environmental densities, as the VAC density
sample is restricted to the range $0.053 < z < 0.093$ to gain an improved
measure of the star formation rate, which is not considered here. We follow the
method of B04: the density is given by $\Sigma_N = N/\pi d_N^2$, where $d_N$ is
the distance to the $N$th nearest neighbour within $\pm 1000 \kms$ in
redshift. This is used to minimize contamination from interlopers. Here the
value of $N$ used is 5, following B04 who choose this value to approximate
\citet{dressler:morphdensity} who uses $N=10$ before correction for
superimposed galaxies. We bin the sample in steps of 0.5 over the range $-1.5 <
\logsigma < 1.5$, a factor of 1,000 in density.

We set a limiting absolute magnitude for galaxies to be counted as neighbours of
$\mrlogh < -19.5$, corresponding to one magnitude fainter than the value of
$M^*$ found for the overall Schechter fit to the galaxy luminosity function of
B06. The corresponding limiting redshift, at which a galaxy of this magnitude
has an apparent magnitude $r = 17.77$ is $z < 0.0889$. This gives a sample
volume-limited to a distance comparable to the $z < 0.08$ of B04. We also
require $z > 0.001$ to exclude very local objects. We measure densities for
galaxies in the range $M^{*+3}_{-2}$, i.e. $-22.5 < \mrlogh < -17.5$, binning in
steps of one magnitude. This again is similar to the range of $-22.23 < \mrlogh
< -17.23$ of B04.

\subsection{Galaxy Colour and Morphology} \label{subsec: colour} 

We use $u-r$ colour following B04 and \citet{baldry:bimodalcmd}, who find that
the galaxy population is well described as the sum of two Gaussians in $u-r$,
with an optimal colour separator of 2.22 \citep{strateva:colorsep}. We use the
model colours from the same dataset as B06, extinction- and K-corrected, as in
that paper, to a rest-frame band-shift of $z=0.1$, denoted throughout in the
plots by $(u-r)_{0.1}$, and in the text by $u-r$. The model colours are derived
from the SDSS model magnitudes. For each galaxy these are derived from the best
fitting of a de Vaucouleurs \citep{devaucouleurs:devprofile} or exponential
profile \citep{freeman:expprofile}.

For morphology, we use three measures: the inverse concentration index,
$\ciinv$, the S\'ersic index, $n$ and the Hubble type. The $\ciinv$ and S\'ersic
index provide a measure of morphology that, while simple, is available for the
full $r < 17.77$ sample. $\ciinv$ is directly from the SDSS DR4 and is given by
$R_{50} / R_{90}$ where $R_N$ is the $N\%$ light Petrosian radius \citep[see,
e.g.][]{stoughton:edr}. The inverse is used because it has the range 0--1. The
S\'ersic index, from the VAGC, is the exponent in the formula for a generalised
galaxy light profile given by \citet{sersic:australes}, or more recently in
e.g. \citet{graham:sersic}. Because $n$ is an exponent, and the bimodality shows
more clearly when it is done so, $n$ is binned logarithmically, as with the
density. As with the colours, these are binned appropriately, giving bins of
$0.23 < \ciinv < 0.58$ in steps of 0.07 and $-0.25 < \logn < 0.75$ in steps of
0.2. We truncate the S\'ersic index at $\logn = 0.75$ to avoid a spike in the
number of galaxies in the VAGC data with $\logn \sim 0.77$. Their data is also
truncated at $n = 6$ ($\logn = 0.778$). Parameters other than colours are
measured in the $r$ band, since this band is used to define the aperture through
which Petrosian flux is measured for all five bands.

The ANN morphologies used were also the same as those in B06, assigned as in
\citet{ball:ann} and updated as described in B06. The types assigned correspond
to the Hubble types E=0, S0=1, Sa=2, Sb=3, Sc=4, Sd=5 and Im=6. As described in
B06, there is bias away from the ends of the scale, particularly for late
types. This is probably due to the dominant contribution of the concentration
index in the training set, which shows a similar bias, and the relative lack of
types of Sd or later in the training set, about 5\% of the total
\citep{fukugita:morph} in part due to the $r$ band selection of the SDSS instead
of the more commonly used bluer bands. Hence the latest types assigned have
$\tann \sim 4.5$. Fig. \ref{fig: galaxies} shows a random sample of five
galaxies from each T type, rounded to the integers 0--4. We bin the ANN
morphologies in five bins from $-0.5$--$4.5$. The colours are also binned in
five bins. Unlike the other measures presented, the limiting magnitude for the
ANN morphologies is $r < 15.9$, as this was the faintest level to which the
visually assigned training set was generated. Of the galaxies studied here,
13,655 have a value for $\tann$ and a valid density. Combined with the absolute
magnitude limit of $\mrlogh < -17.5$, these limits ensure that there is no
extrapolation from the ANN training set in apparent or absolute magnitude.

For each set of measurements, the galaxies are required to have values within
the ranges of the bins of all quantities that are binned, i.e. density,
luminosity, and one or both of colour and morphology. This ensures that each
related figure is showing the same galaxies. The resulting samples consist of
79,553 galaxies for $u-r$, 79,606 for $\ciinv$, 75,861 for S\'ersic index and
13,655 for $\tann$. For the residuals in $u-r$ and morphology, requiring both
colour and morphology to be within the bin ranges reduces the sample sizes to
79,495, 75,753 and 13,638 respectively.

\begin{figure*}
\centering
\includegraphics[width=0.19\textwidth]{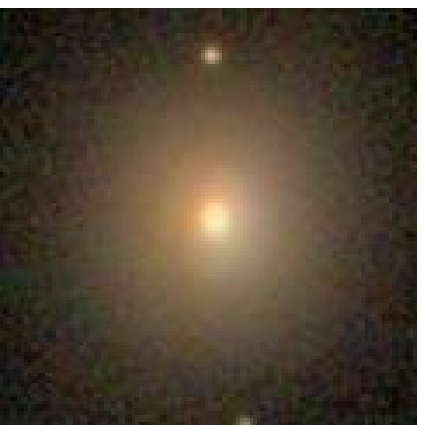}
\includegraphics[width=0.19\textwidth]{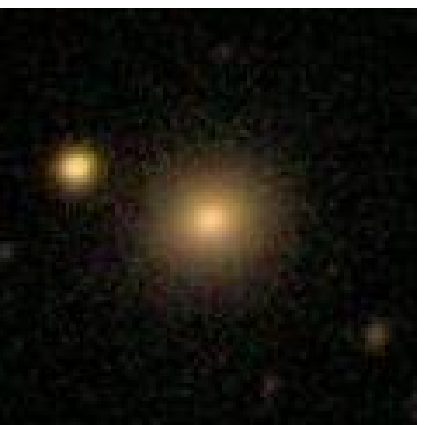}
\includegraphics[width=0.19\textwidth]{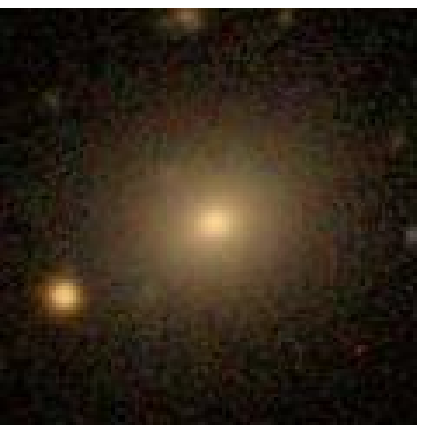}
\includegraphics[width=0.19\textwidth]{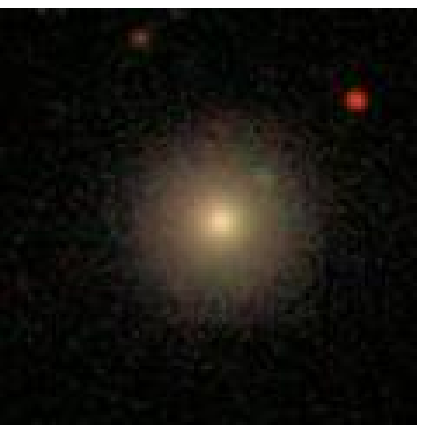}
\includegraphics[width=0.19\textwidth]{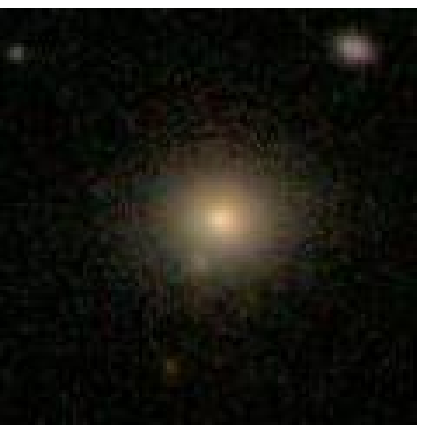}
\includegraphics[width=0.19\textwidth]{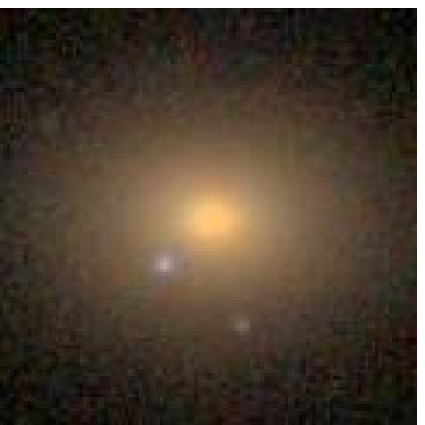}
\includegraphics[width=0.19\textwidth]{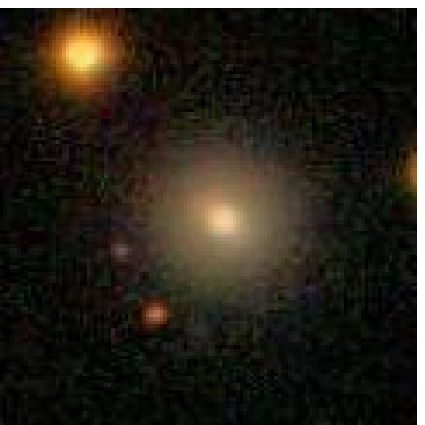}
\includegraphics[width=0.19\textwidth]{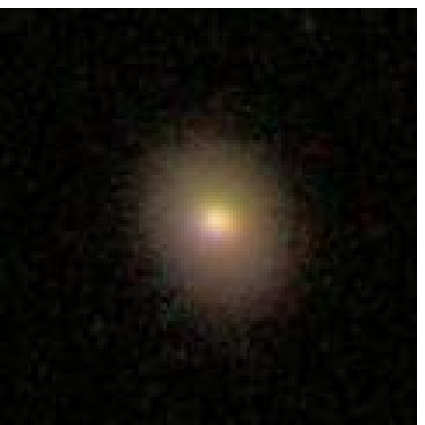}
\includegraphics[width=0.19\textwidth]{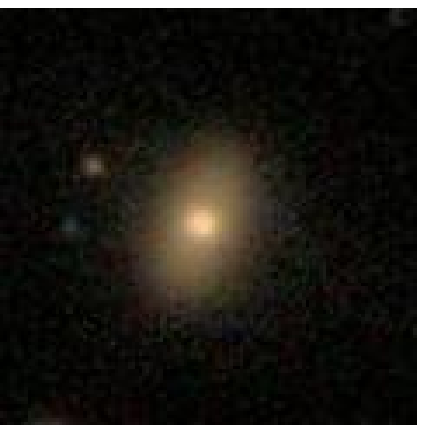}
\includegraphics[width=0.19\textwidth]{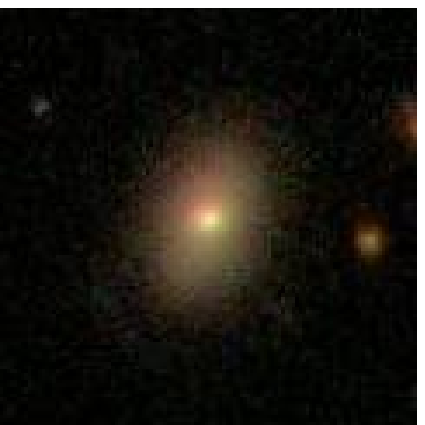}
\includegraphics[width=0.19\textwidth]{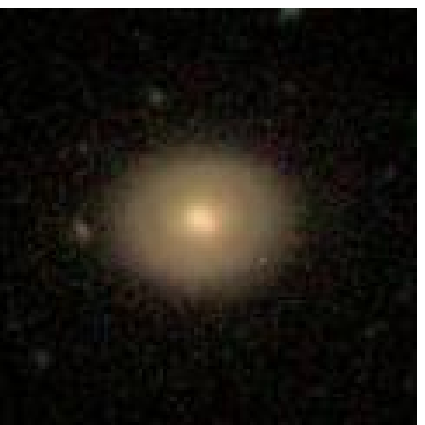}
\includegraphics[width=0.19\textwidth]{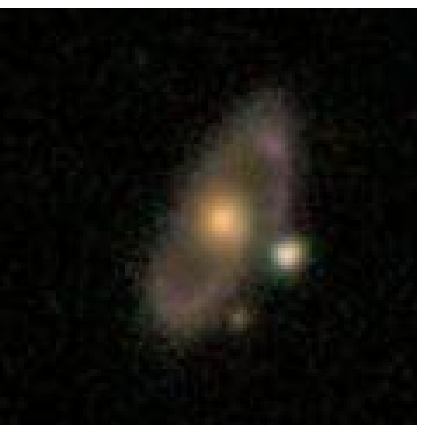}
\includegraphics[width=0.19\textwidth]{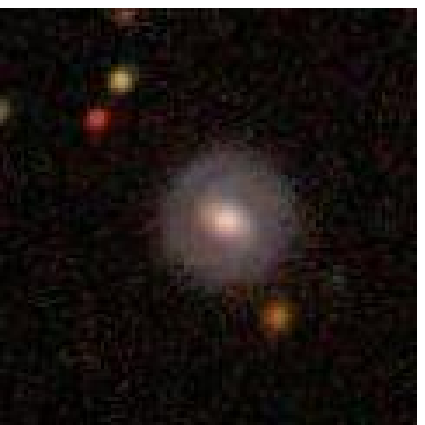}
\includegraphics[width=0.19\textwidth]{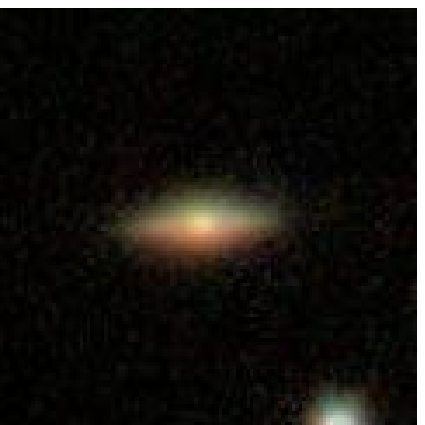}
\includegraphics[width=0.19\textwidth]{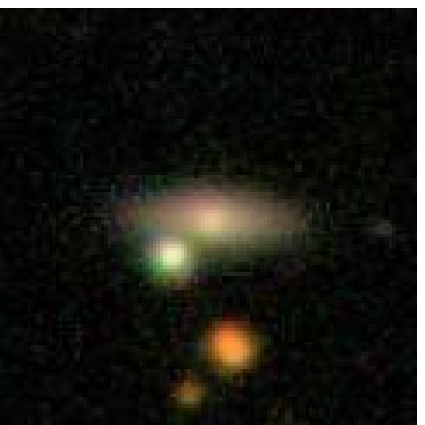}
\includegraphics[width=0.19\textwidth]{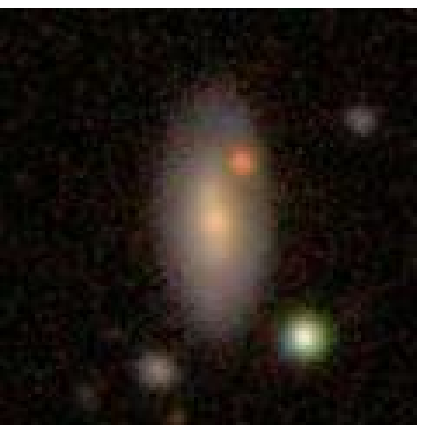}
\includegraphics[width=0.19\textwidth]{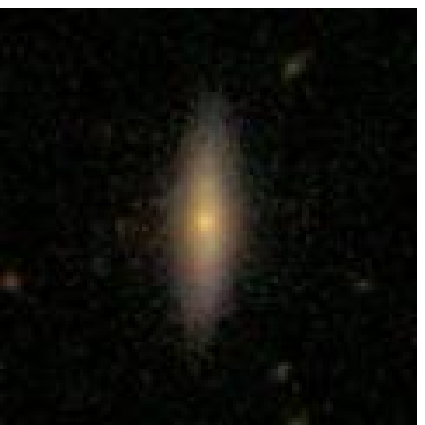}
\includegraphics[width=0.19\textwidth]{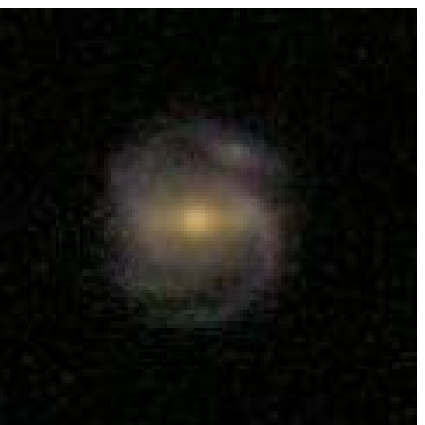}
\includegraphics[width=0.19\textwidth]{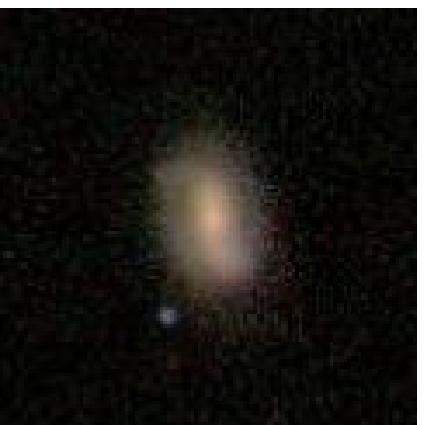}
\includegraphics[width=0.19\textwidth]{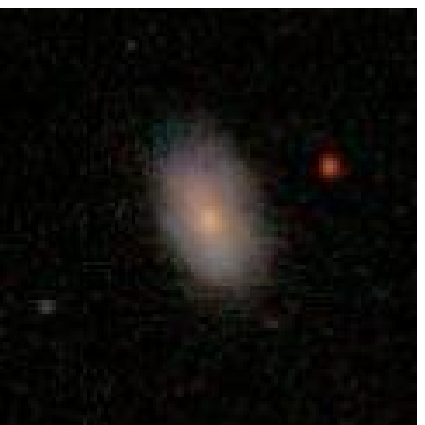}
\includegraphics[width=0.19\textwidth]{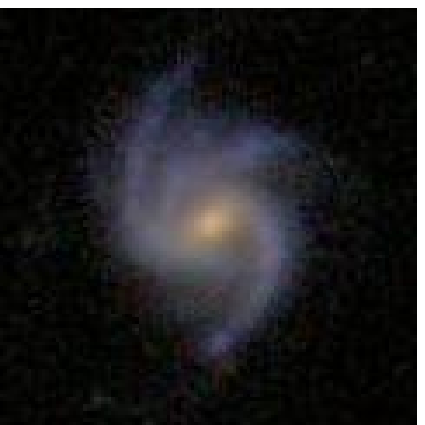}
\includegraphics[width=0.19\textwidth]{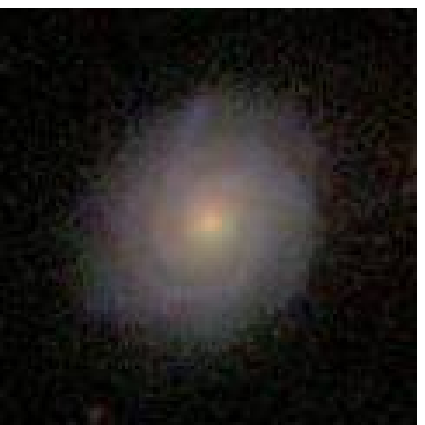}
\includegraphics[width=0.19\textwidth]{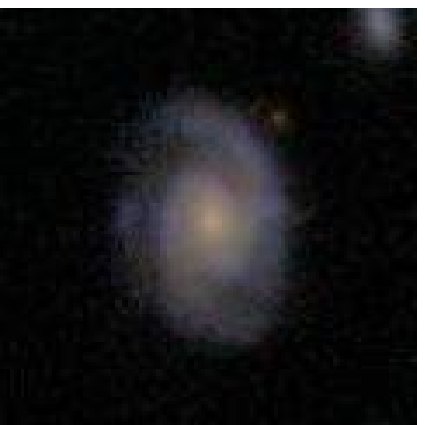}
\includegraphics[width=0.19\textwidth]{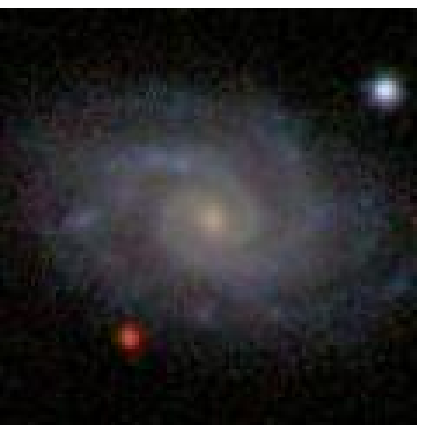}
\includegraphics[width=0.19\textwidth]{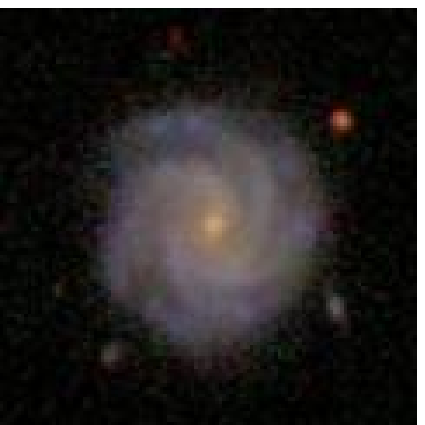}
\caption{SDSS cutout images showing five random galaxies of each of the T-types
  0--4 (rounded to the nearest integer) on respective rows. The T types
  correspond to Hubble types E, S0, Sa, Sb and Sc. This shows the typical
  quality of the data used in determining the SDSS object parameters used in
  training the artificial neural network to assign the types. A larger selection
  of similar images can be found in \citet{fukugita:morph}. \label{fig:
    galaxies}}
\end{figure*}

\subsection{Data Fitting} \label{subsec: fitting} 

The data points for the histograms of colour and morphology for bins of
$\logsigma$ and $M_r$ are fitted with a double Gaussian using a simplex search
algorithm over the ranges $0.5 < u-r < 4.0$, $0.23 < \ciinv < 0.58$, $-0.25 < n
< 0.75$ and $0 < \tann < 4.5$. The number of galaxies outside these ranges is
small, being less than 1\% of the total. The fits involve six parameters and are
given by \begin{equation} A~N(\overline x_1,\sigma_1) + B~N(\overline
  x_2,\sigma_2)   , \end{equation} where the $N(\overline x,\sigma)$ are
Gaussians with mean $\overline x$ and standard deviation $\sigma$ and $A$ and
$B$ are scalars. Whereas B04 constrain the $\sigma$ values of the Gaussians to
be functions of $M_r$ only, here we do not apply any restrictions on the
parameters of the two Gaussians in one plot relative to any others. The fits do
not utilise the extreme ends of the distributions, where the number of galaxies
is negligible.

Again following B04, the $1\sigma$ error bars for each bin are given by
$\sqrt{N_{\mrm{bin}}+2}$ where $N_{\mrm{bin}}$ is the number of galaxies in that
bin. This is an approximation of the Poisson error for small number statistics.

\section{Results} \label{sec: results} 

\subsection{Two-Dimensional Distributions} \label{subsec: 2d} 

\begin{figure*} \centering
\includegraphics[width=\textwidth]{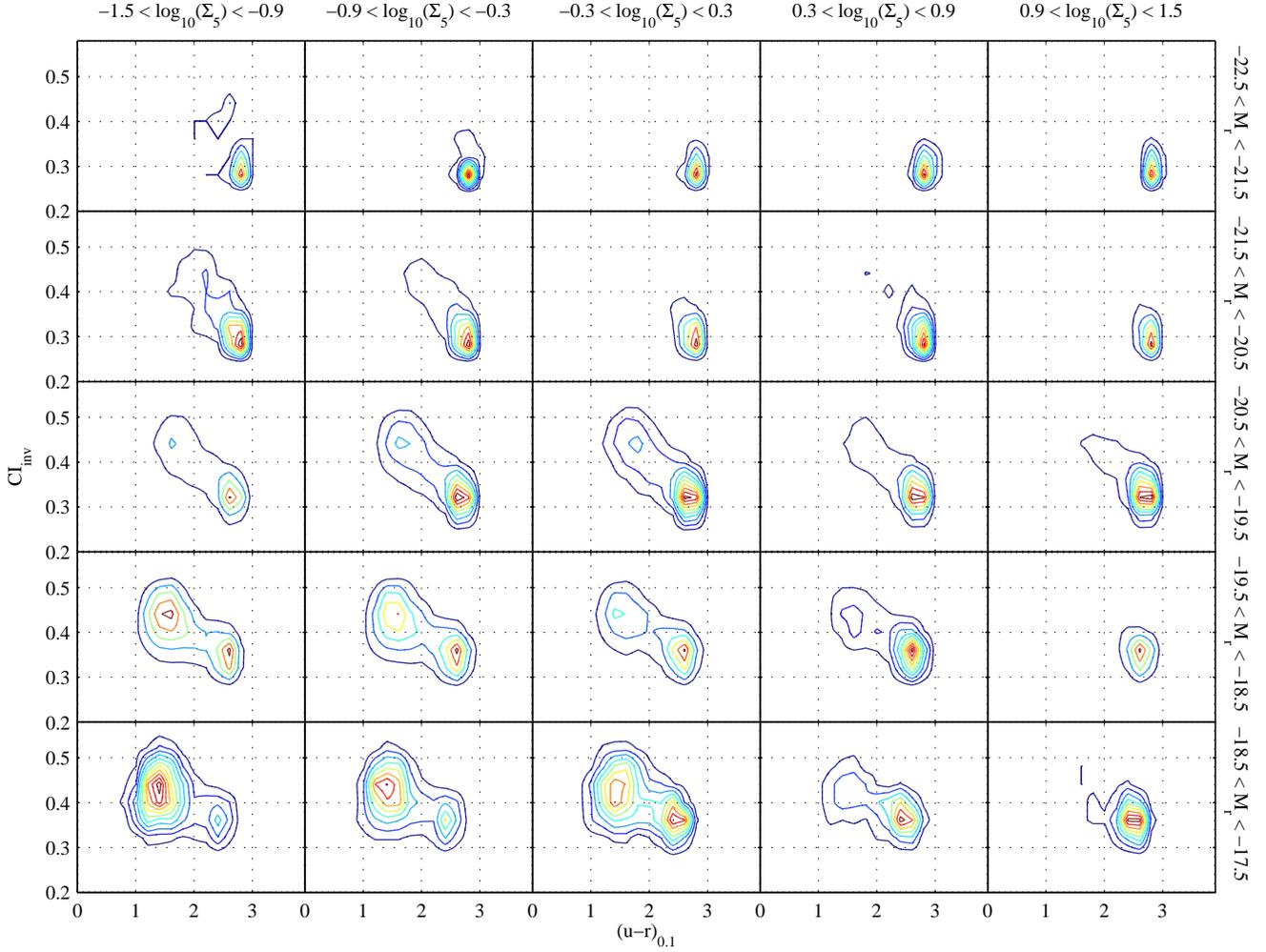}
\caption{Contour plot of Petrosian inverse concentration index, $\ciinv$, versus
  rest-frame $u-r$ colour, subdivided into panels by environmental density,
  $\Sigma_5$ and absolute magnitude, $M_r$. The sample consists of 79,606
  galaxies. The contours are normalised to the numbers of galaxies in each
  panel, so small numbers lie outside the lowest contour. \label{fig: CI
    contour}}
\end{figure*}

\begin{figure*} \centering
\includegraphics[width=\textwidth]{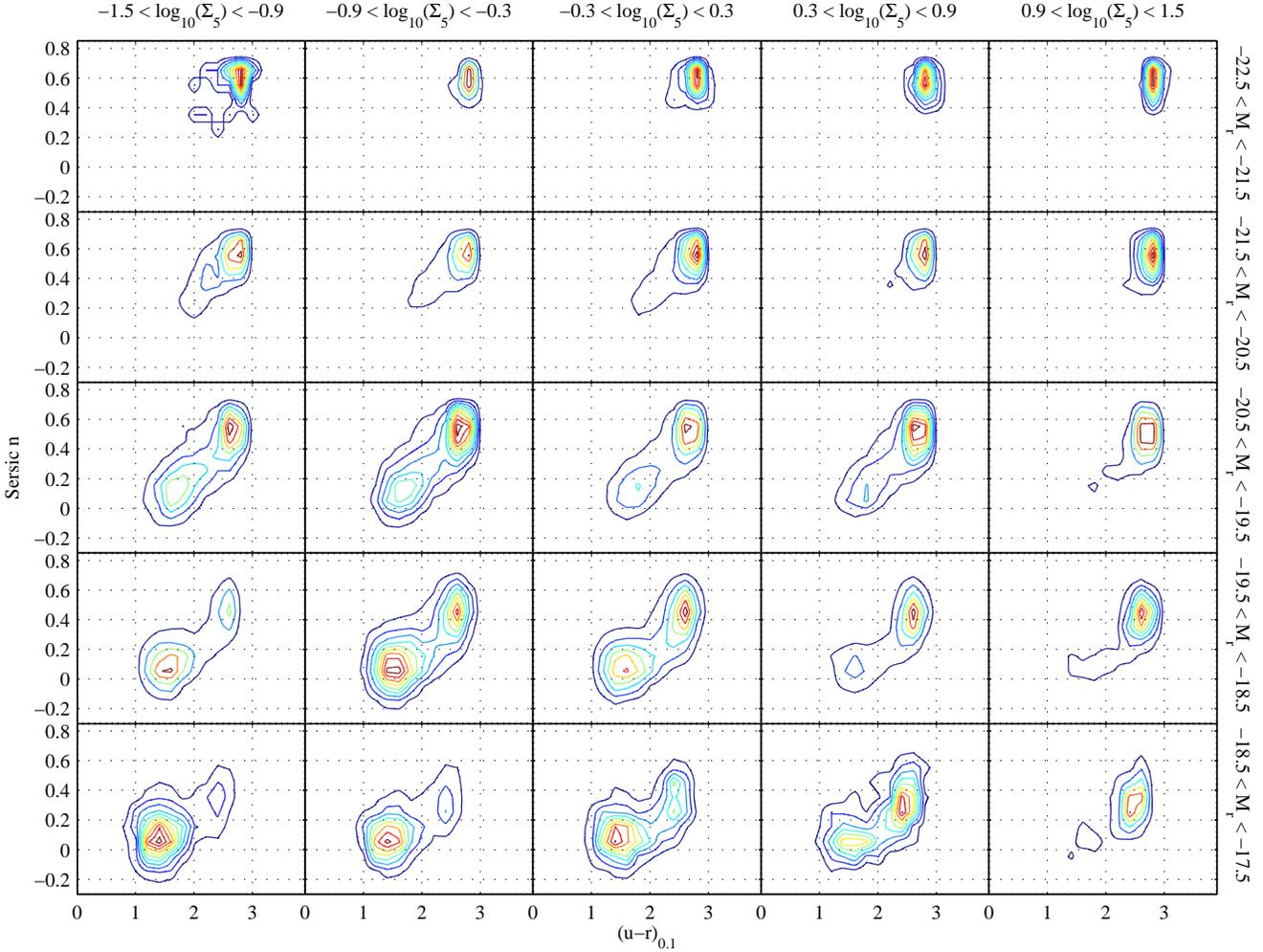}
\caption{As Figure \ref{fig: CI contour}, but with $\ciinv$ replaced by the
  logarithmic S\'ersic index, $\logn$, for 75,861 galaxies. The truncation of
  the data at $\logn=0.75$ (\S \ref{subsec: colour}) is visible. \label{fig:
    Sersic contour}}
\end{figure*}

\begin{figure*} \centering
\includegraphics[width=\textwidth]{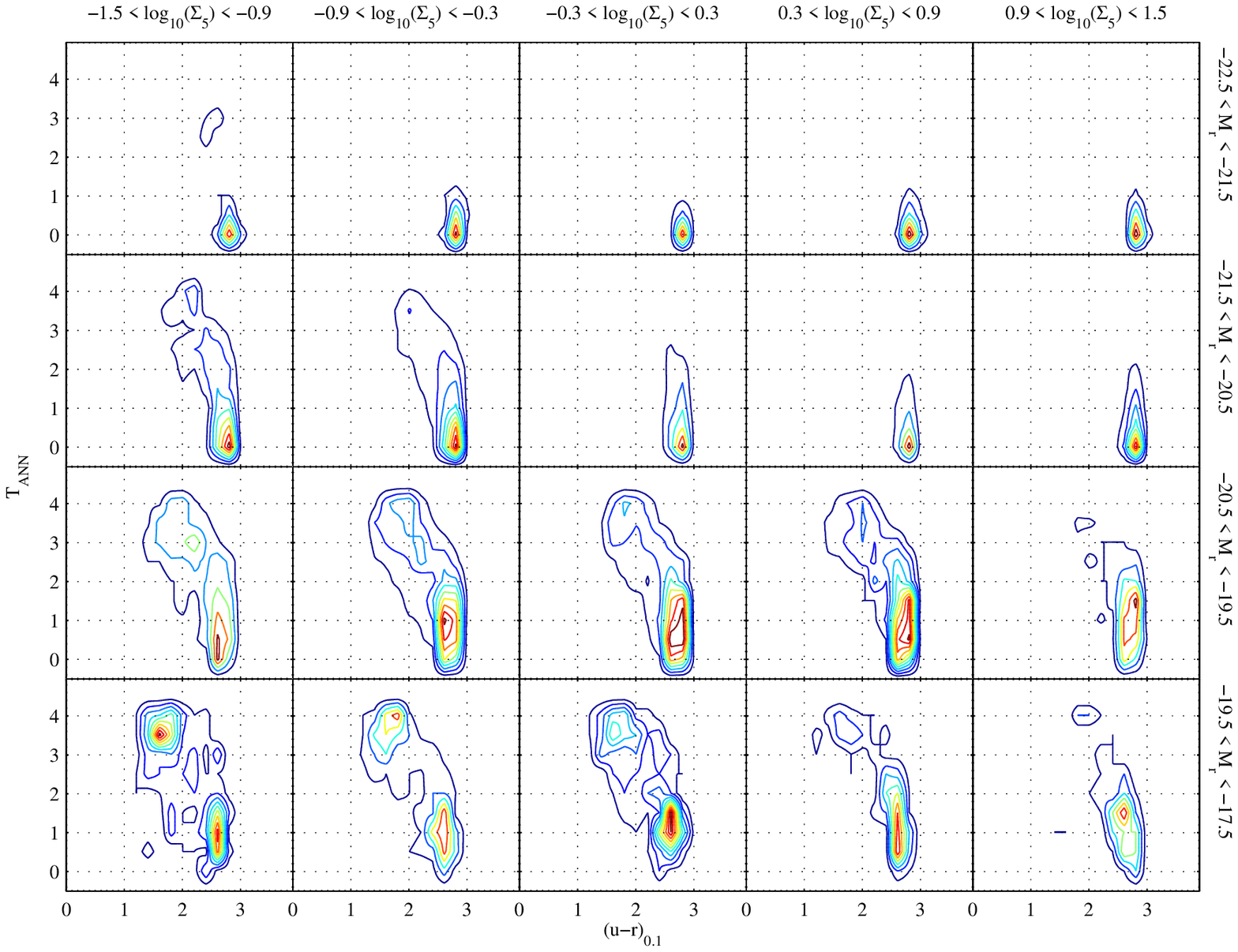}
\caption{As Figure \ref{fig: CI contour}, but with $\ciinv$ replaced by the
  Hubble T type assigned by an artificial neural network, $\tann$, and
  the faintest two bins combined due to the low number of galaxies in the
  faintest bin. The sample size is 13,655 galaxies. \label{fig: T contour}}
\end{figure*}

Figs. \ref{fig: CI contour}--\ref{fig: T contour} show colour versus morphology
($\ciinv$, S\'ersic $n$ and $\tann$ respectively) for the sample, divided into
panels according to density and luminosity. It is clear that there is a broad
division into two populations. This division justifies the treatment below of
the galaxies as two populations modelled by a double Gaussian, which is used in
\S \ref{subsec: gaussian}--\S \ref{subsec: residual} below to quantify the
trends seen in more detail.

In Fig. \ref{fig: CI contour}, the populations are approximately centred on $u-r
\sim 2.75$, $\ciinv \sim 0.35$ and $u-r \sim 1.5$, $\ciinv \sim 0.45$. These
correspond to the red, early-type and blue, late-types. As expected, there are
few red galaxies with high $\ciinv$ or blue galaxies with low $\ciinv$, and
there is little dependence of each population centre on luminosity or
density. What does change radically is the relative numbers of galaxies in each
population, which is emphasized here by the normalisation applied to the
contours in each panel. Blue colours and diffuse morphologies dominate
in galaxies of low luminosity in low density environments. They are
almost entirely absent in luminous galaxies and in high-density
regions.

Fig. \ref{fig: Sersic contour} shows a similar pattern, although the sense of
$n$ is inverted compared to $\ciinv$. The truncation in the S\'ersic index at
$\logn < 0.75$ (\S \ref{subsec: colour}) is visible, although the contours show
that few galaxies are excluded. At fainter magnitudes, $M_r \simgt -19.5$, the
visual overlap between the populations is lower in $n$ than in $\ciinv$.

Fig. \ref{fig: T contour} shows that there are again similar patterns in $\tann$
morphology. Types $0 \simlt \tann \simlt 2$, i.e., E--Sa, are all approximately
the same colour, although the incidence of types $\tann \simgt 1$ decreases at
bright magnitudes $M_r \simlt -20.5$. There are relatively few galaxies at the
faint end, $-17.5 < M_r < -18.5$ due to the brighter $r < 15.9$ magnitude limit
for $\tann$ compared to $r < 17.77$ for $\ciinv$ and $n$.

\subsection{Gaussian Fits} \label{subsec: gaussian} 

\begin{figure*} \centering
\includegraphics[width=\textwidth]{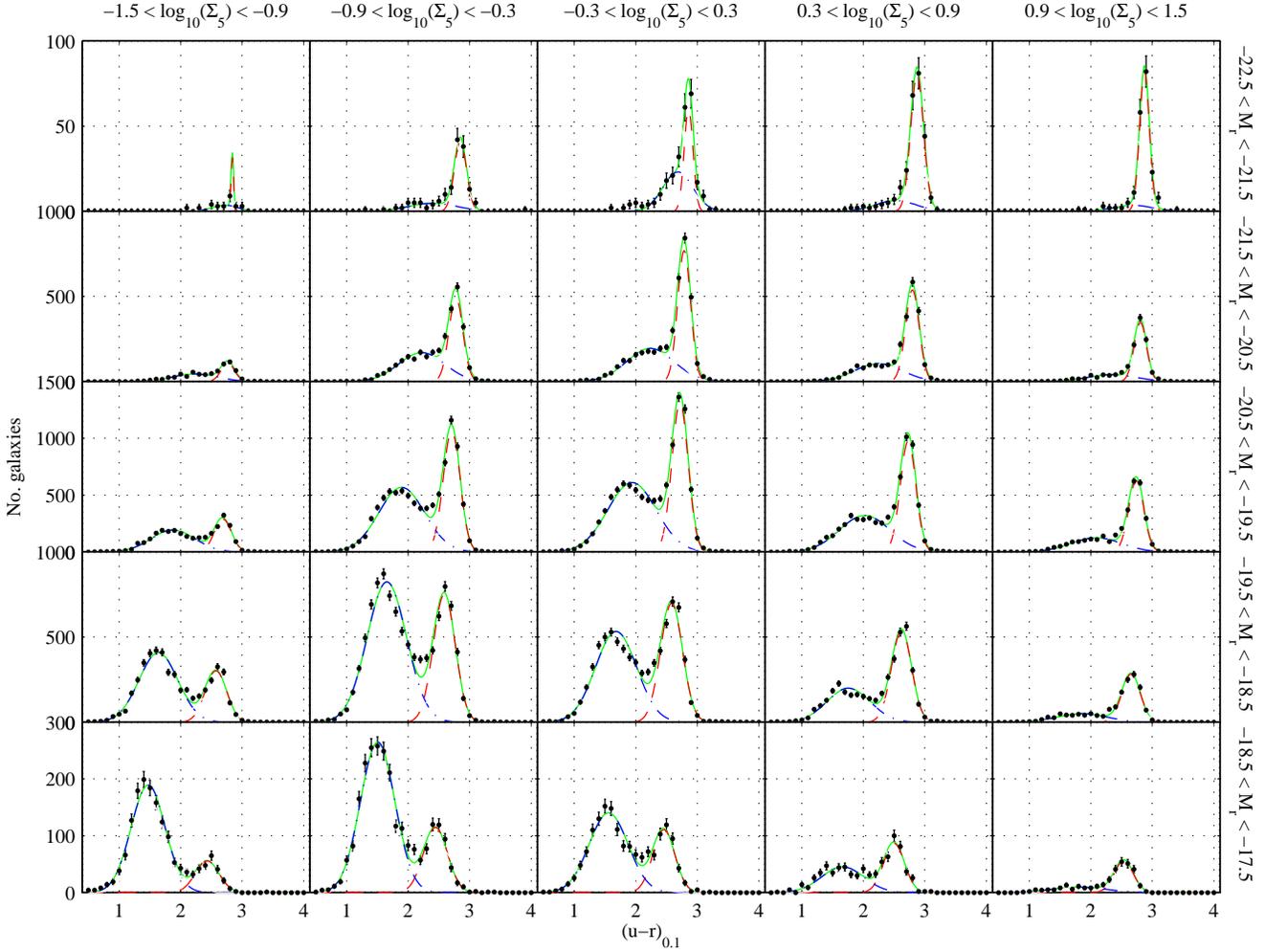}
\caption{Numbers of galaxies for $u-r$ subdivided by $\Sigma_5$ and $M_r$. The
  points are the number of galaxies in each bin; the error bars are given by
  $\sqrt{N_{\mrm{bin}}+2}$ where $N_{\mrm{bin}}$ is the number of galaxies in
  that bin; the solid line is the best fit of the sum of two Gaussians; the
  dot-dash and dashed lines are the individual Gaussians. The vertical axes
  within each row of panels are given equal heights for clarity. This plot is
  similar to that of \citet{balogh:bimodallfenvt}, but for DR4 as opposed to
  DR1. The trends seen are very similar, as expected, because DR4 is a superset
  of DR1. \label{fig: CDL}}
\end{figure*}

Fig. \ref{fig: CDL} shows the rest-frame $u-r$ colour distribution, subdivided
by density $\logsigma$ and luminosity $M_r$. The distributions are clearly
bimodal and well fit by a sum of two Gaussians, confirming the idea of two
separate populations described by numerous authors
\citep[e.g.][]{baldry:bimodalcmd}. As expected, the fraction of blue galaxies
decreases with density. The position of the blue peak moves redwards from $u-r
\approx 1.5$ to $u-r \approx 2.25$ and that of the red peak stays approximately
constant, with just a slight reddening from $u-r \approx 2.5$ to $u-r \approx
2.75$. Unlike the peaks, the position of the minimum between the two
distributions stays constant, although in the higher density and luminosity bins
the exact position of the minimum is unclear. Versus luminosity, a similar trend
is seen with the red peak moving slightly bluewards and dropping in fraction
with decreasing luminosity and the blue peak becoming bluer. Bright galaxies at
$\mrlogh < -21.5$ are relatively rare as are faint galaxies due to the $r <
17.77$ limit for spectra.

The figure represents an update of fig. 1 of B04, using DR4 rather than DR1 and
with the differences described in \S\ref{sec: data} above. As mentioned, B04
find that at fixed luminosity the mean colours of the two distributions are
nearly independent of density, with just a weak trend for colours to become
redder at high densities, but that the fraction of red galaxies rises sharply at
all luminosities. As expected, our plot shows the same trends. 

\begin{figure*} \centering
\includegraphics[width=\textwidth]{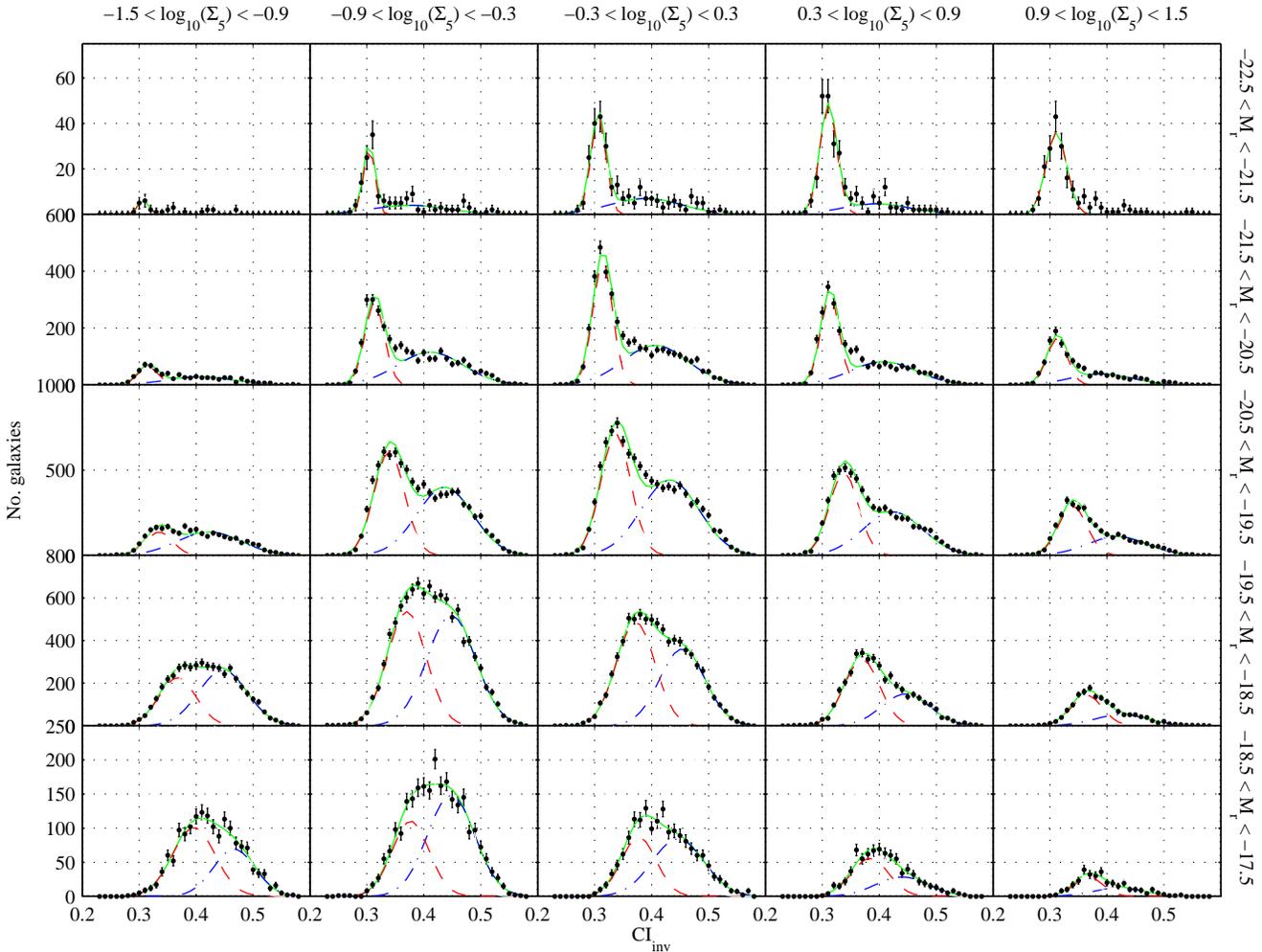}
\caption{As Fig. \ref{fig: CDL} but for $\ciinv$. \label{fig: MDLci}}
\end{figure*}

\begin{figure*} \centering
\includegraphics[width=\textwidth]{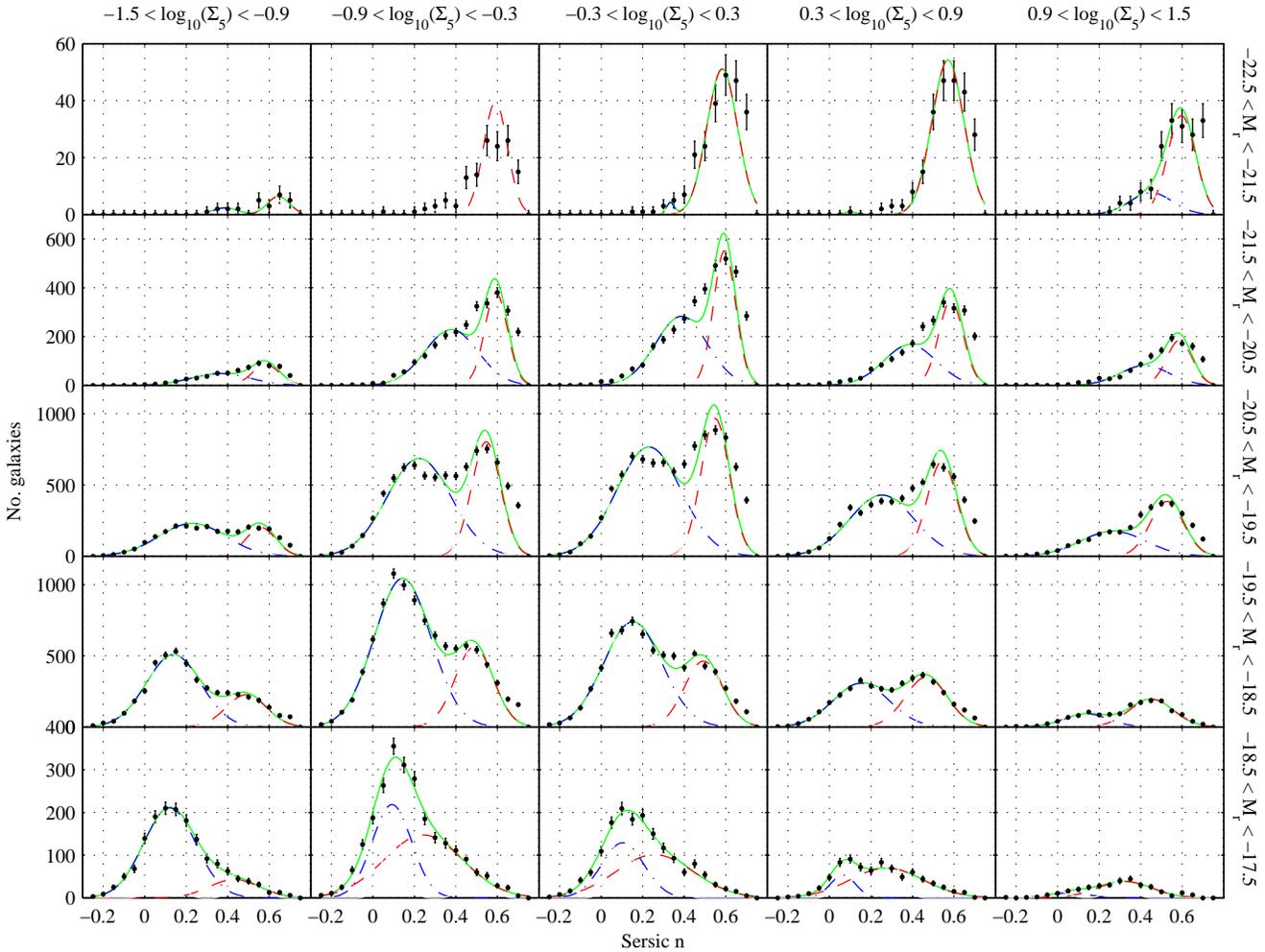}
\caption{As Fig. \ref{fig: CDL} but for S\'ersic $n$. \label{fig: MDLn}}
\end{figure*}

\begin{figure*} \centering
\includegraphics[width=\textwidth]{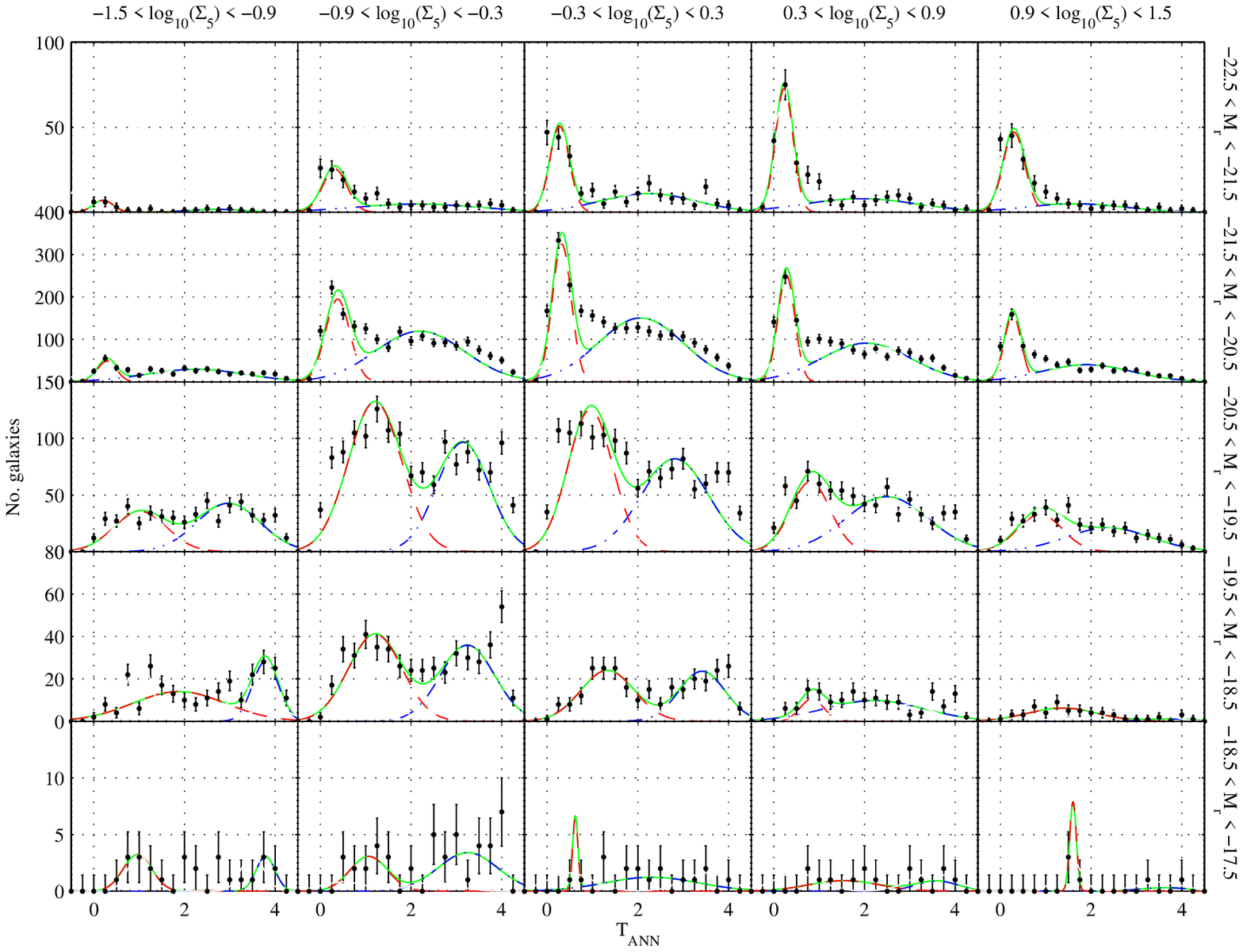}
\caption{As Fig. \ref{fig: CDL} but for $\tann$. \label{fig: MDLt}}
\end{figure*}

Figs. \ref{fig: MDLci}--\ref{fig: MDLt} show the same as Fig. \ref{fig: CDL} but
for the $\ciinv$, S\'ersic $n$ and $\tann$ morphologies respectively, i.e., they
show the morphology-density relation subdivided by luminosity. 

As with colour, the concentration index shows clearly bimodal distribution,
although significant overlap between the populations causes a unimodal
appearance in some of the panels. The plots show peaks at $\ciinv \sim 0.3$ and
$\ciinv \sim 0.43$, corresponding to the de Vaucouleurs elliptical galaxy
profile and the exponential spiral disc profiles, respectively. As expected,
early-type galaxies dominate at high density and high luminosity.

The S\'ersic index shows a bimodal distribution at all densities and all but the
highest and lowest luminosities. The indices $n = 4$ ($\logn = 0.60$) and $n =
1$ ($\logn = 0$) correspond to the de Vaucouleurs and exponential profiles. The
peaks in the distribution are close to these values, but slightly nearer the
overall mean, in a similar manner to $\ciinv$.

For $\tann$, a sum of two Gaussians is now a worse fit to the data, with a clear
excess of galaxies at $\tann \sim 1$ at densities $\logsigma \simgt -0.3$. The
expected dominance of early types at high density is seen and the peak narrows
with density. The late-type peaks becomes lower and earlier, in a similar manner
to the reddening of the blue peak in $u-r$. The lack of (or less clear) division
into a bimodal population suggests that if galaxies really are divided into just
two dominant populations then either (1) the $\tann$ is insensitive to it,
spreading between the two; (2) that $\tann$ is more sensitive to transitional
states between the two, such as S0 galaxies, than colour, or (3) the $\tann$ is
an intrinsically `fuzzy' measure, resulting in a inevitable spread across
types. If (2) is correct, such a transitional state must then be rare and/or
brief in colour space for the clear division in colour to be seen as S0 galaxies
are not rare. That the transitional population is clearest at higher densities
is also consistent with it being S0s. (3) is less likely for the overall
population, i.e. one can split E/S0 from spiral, but the excess over Gaussian at
$\tann \sim 1$ and at very late types supports the `fuzzy' nature of the Hubble
type classification. The spread is not likely to be an artefact of typing via
neural network, as these were shown \citep[e.g.][]{storrielombardi:ann,
  naim:annmorph, ball:ann} to perform with an RMS spread between true type and
assigned type no different to that between human classifiers.

In most of the plots in Figs. \ref{fig: CDL}--\ref{fig: MDLt}, the reduced
$\chi^2$ values, expected to be around 1 or less, of the double Gaussian fit are
large. This was also seen in the $\chi^2$ for fitting the bivariate LF in
B06. As with those results, this is probably due to the small size of the error
bars resulting from the large number of galaxies in the sample, indicating that
the fitting function used is only an approximation to the data. What is clear is
that the data is better fitted by two Gaussians than, say, one.

\subsection{Red Fraction, Mean, Dispersion} \label{subsec: fraction} 

\begin{figure*} \centering
\mbox{\subfigure{\includegraphics[width=3in]{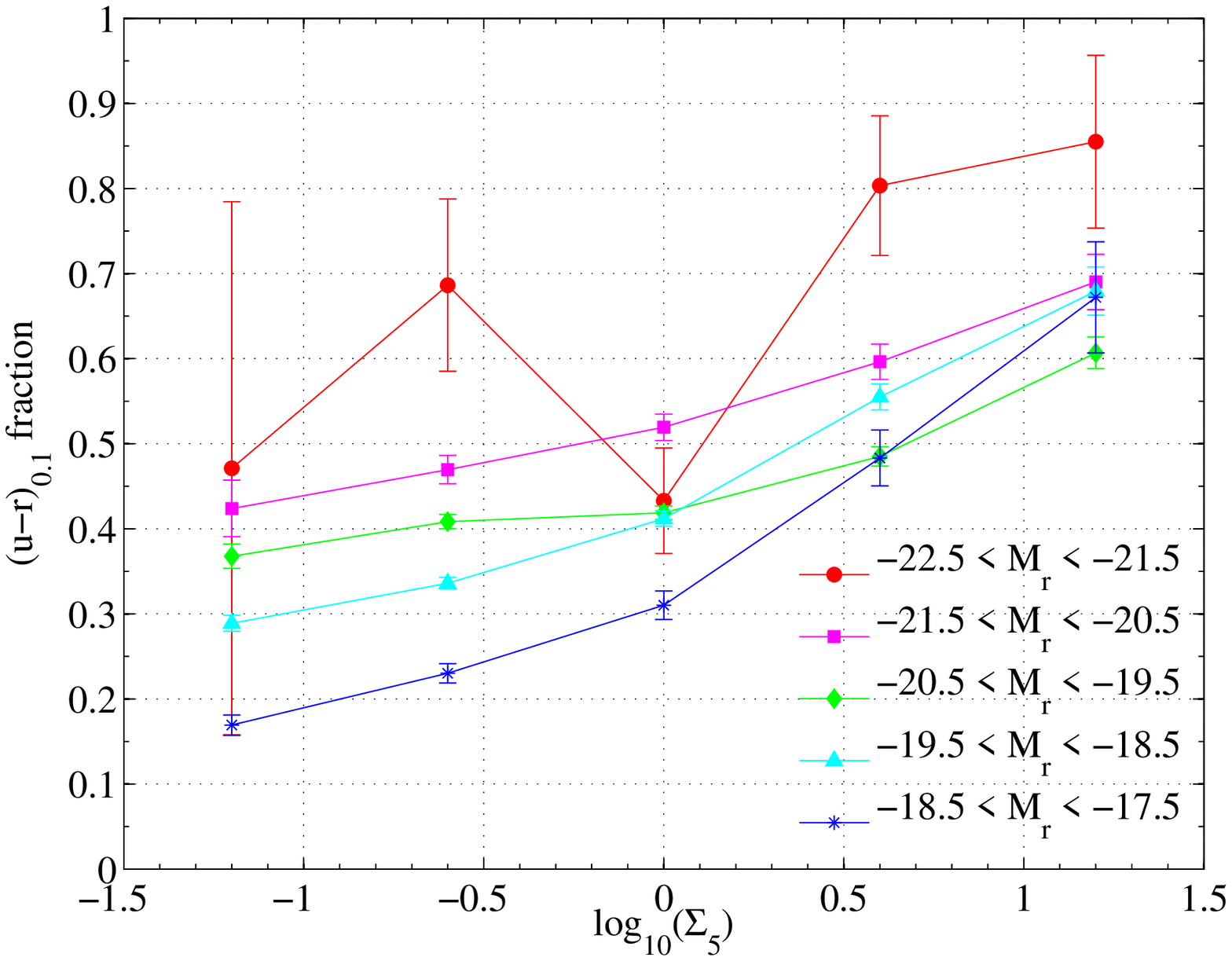}}\qquad
      \subfigure{\includegraphics[width=3in]{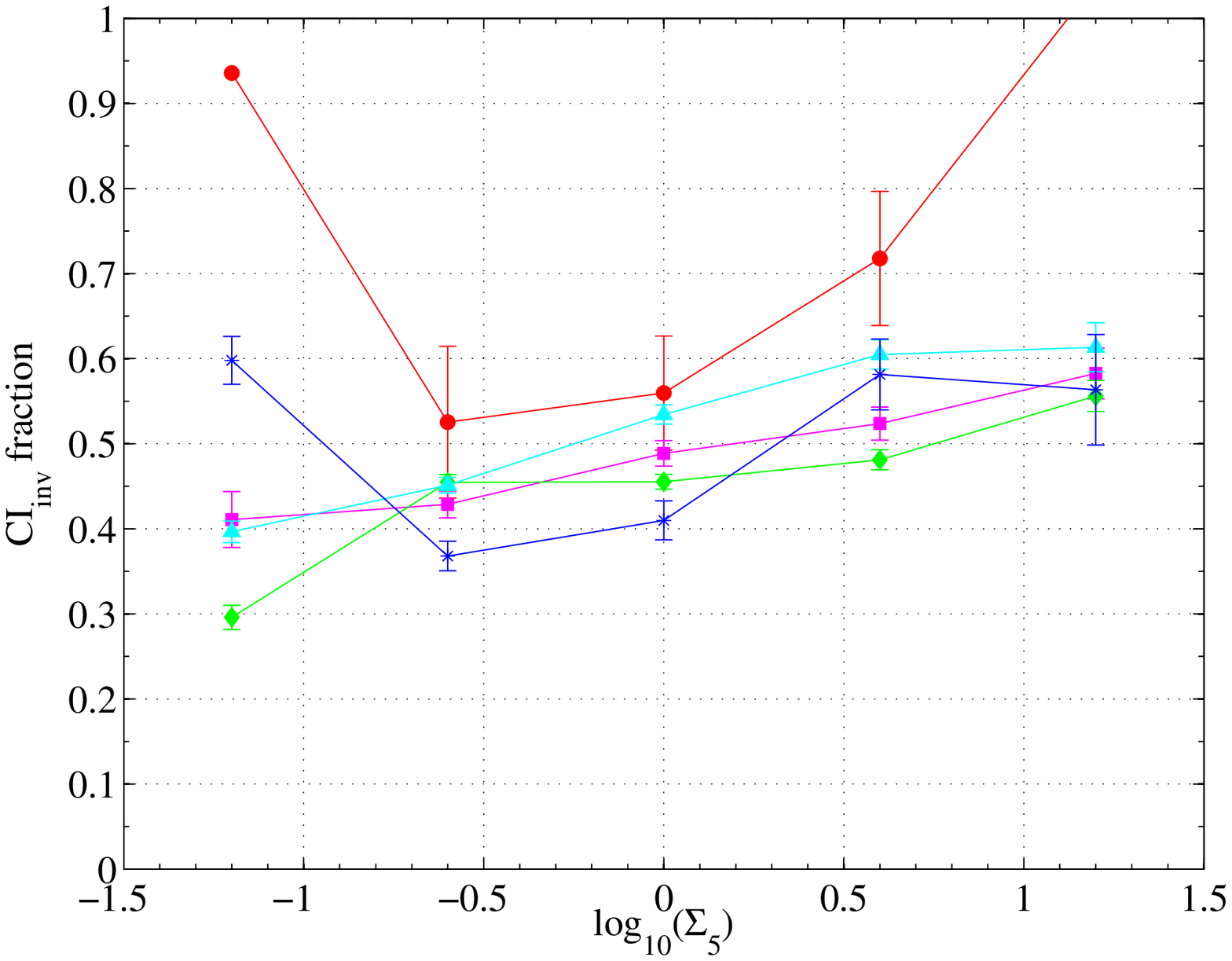}}}
\mbox{\subfigure{\includegraphics[width=3in]{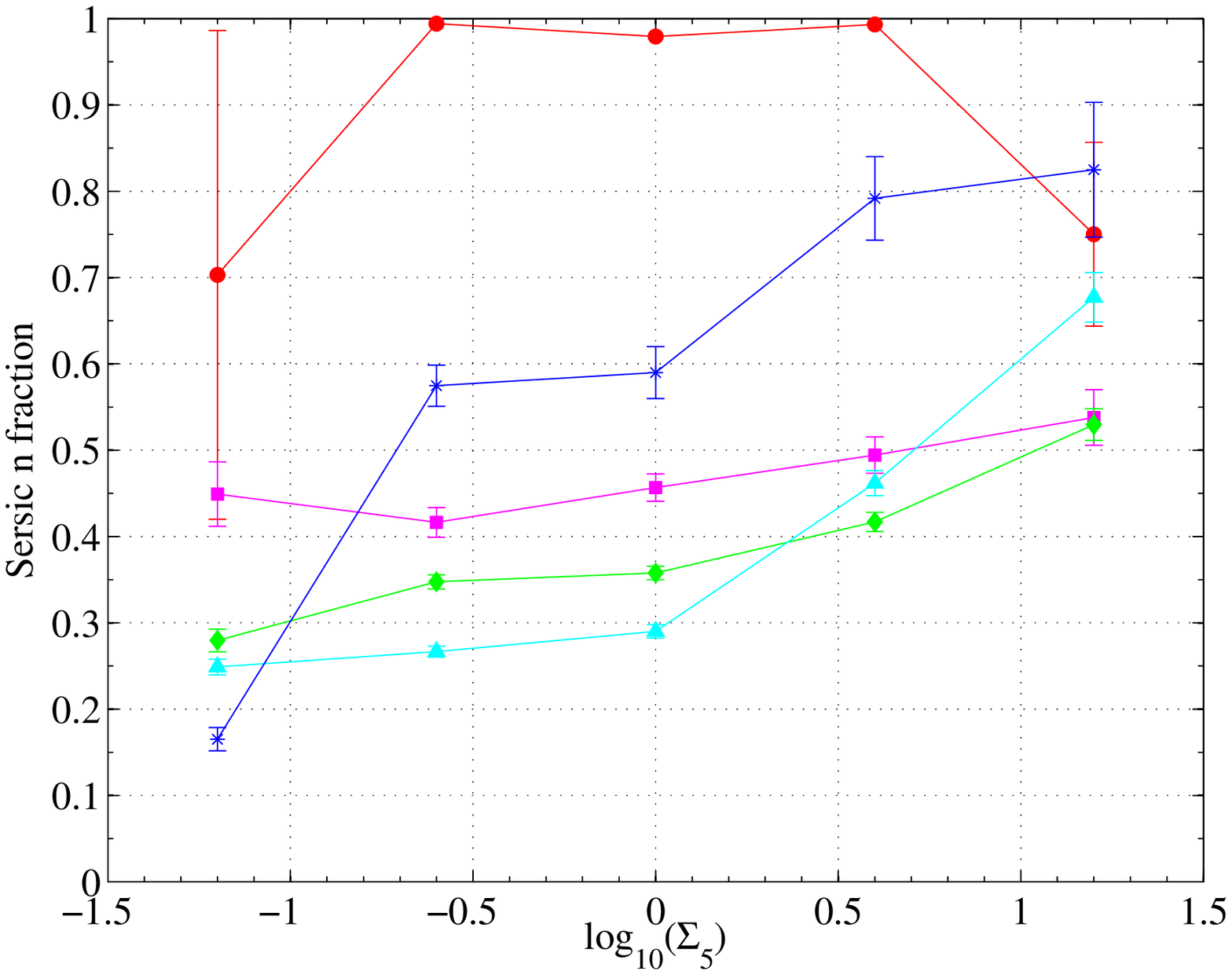}}\qquad
      \subfigure{\includegraphics[width=3in]{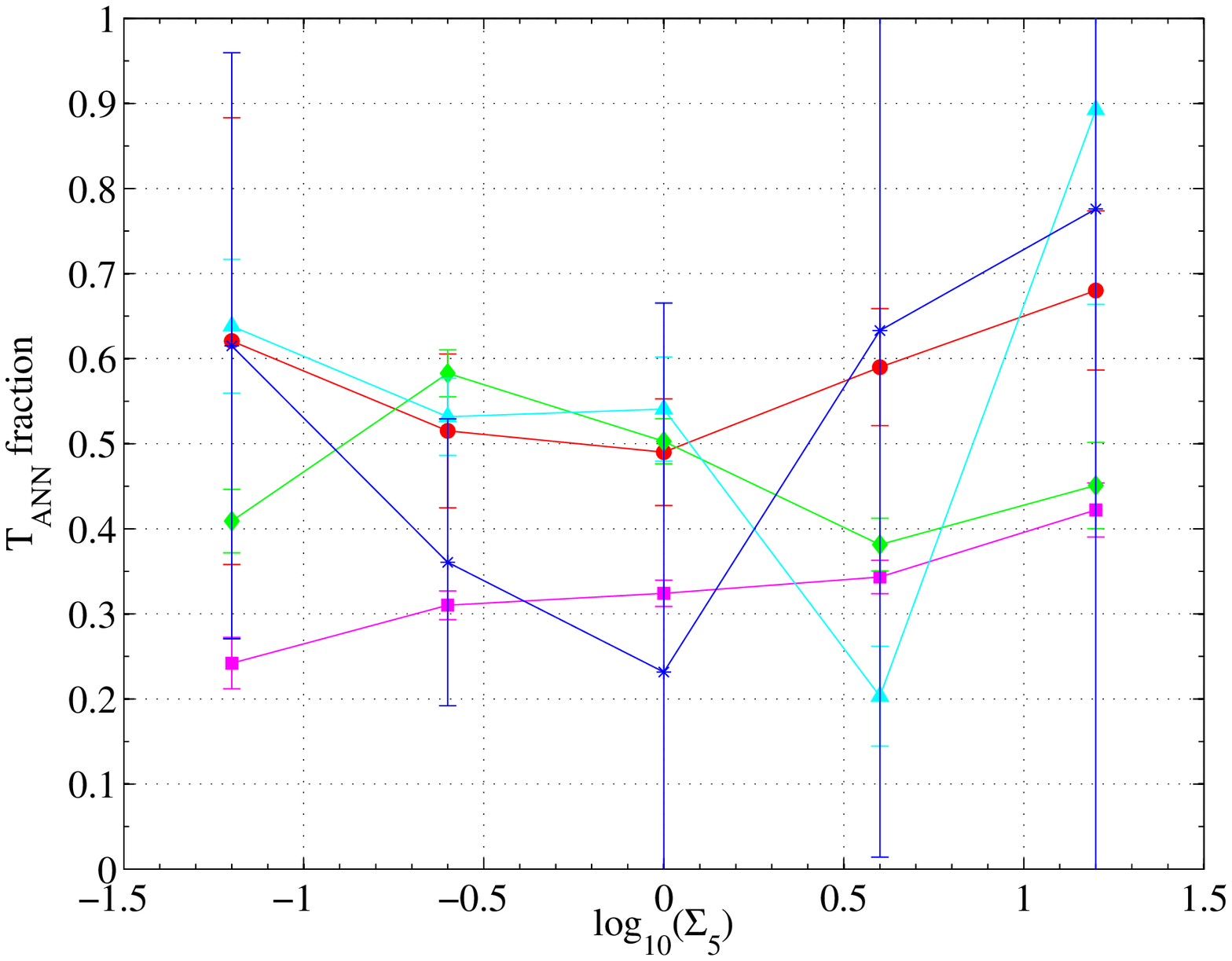}}}
\caption{Red and early-type galaxy fractions in each $M_r$ bin versus $\Sigma_5$
  for $u-r$, $\ciinv$, S\'ersic $n$ and $\tann$. The error bars are
  $1\sigma$. The left hand panel shows a similar upward trend to fig. 2 of
  B04. \label{fig: Red fractions}}
\end{figure*}

\begin{figure*} \centering
\mbox{\subfigure{\includegraphics[width=3in]{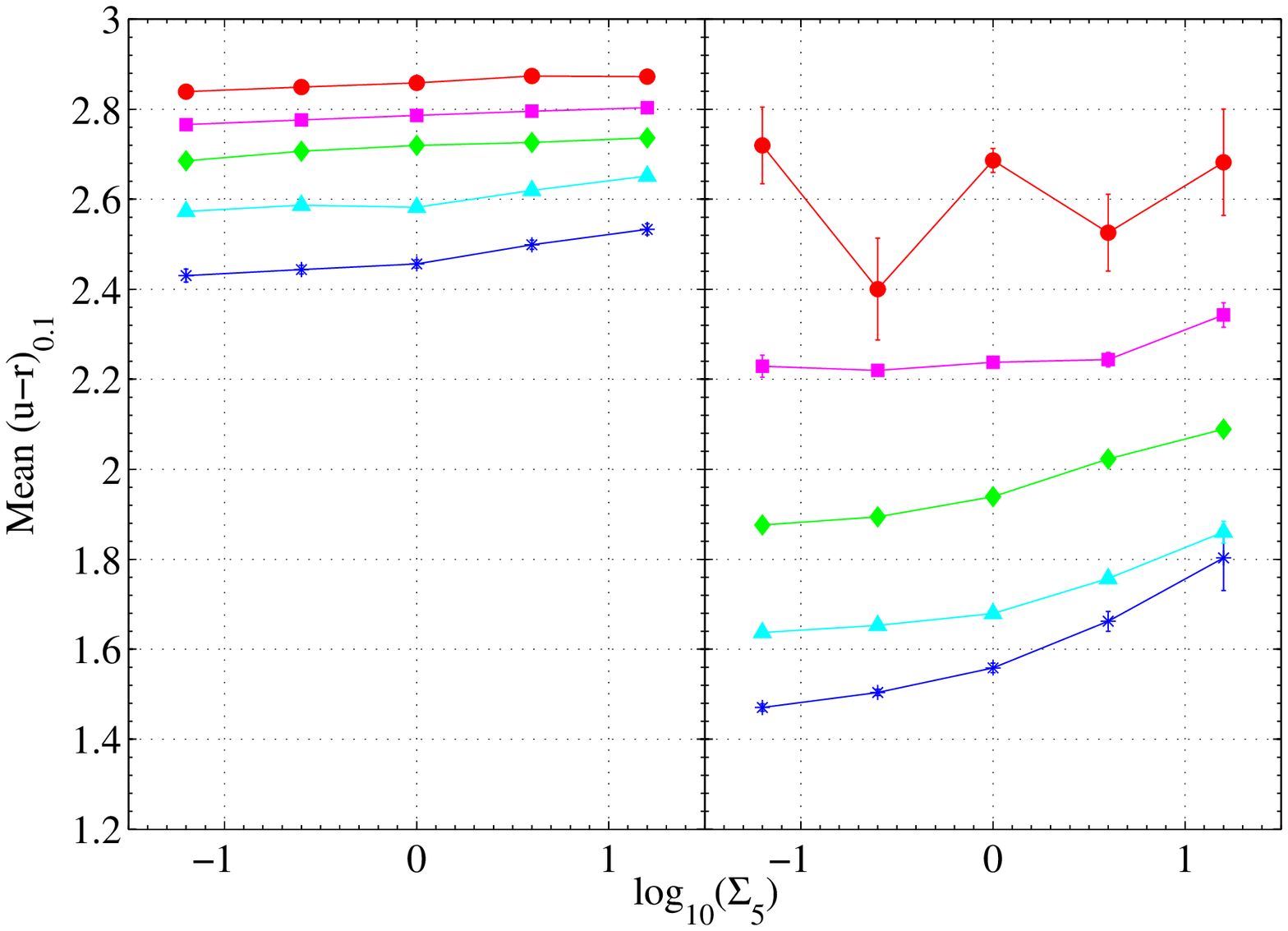}}\qquad
      \subfigure{\includegraphics[width=3in]{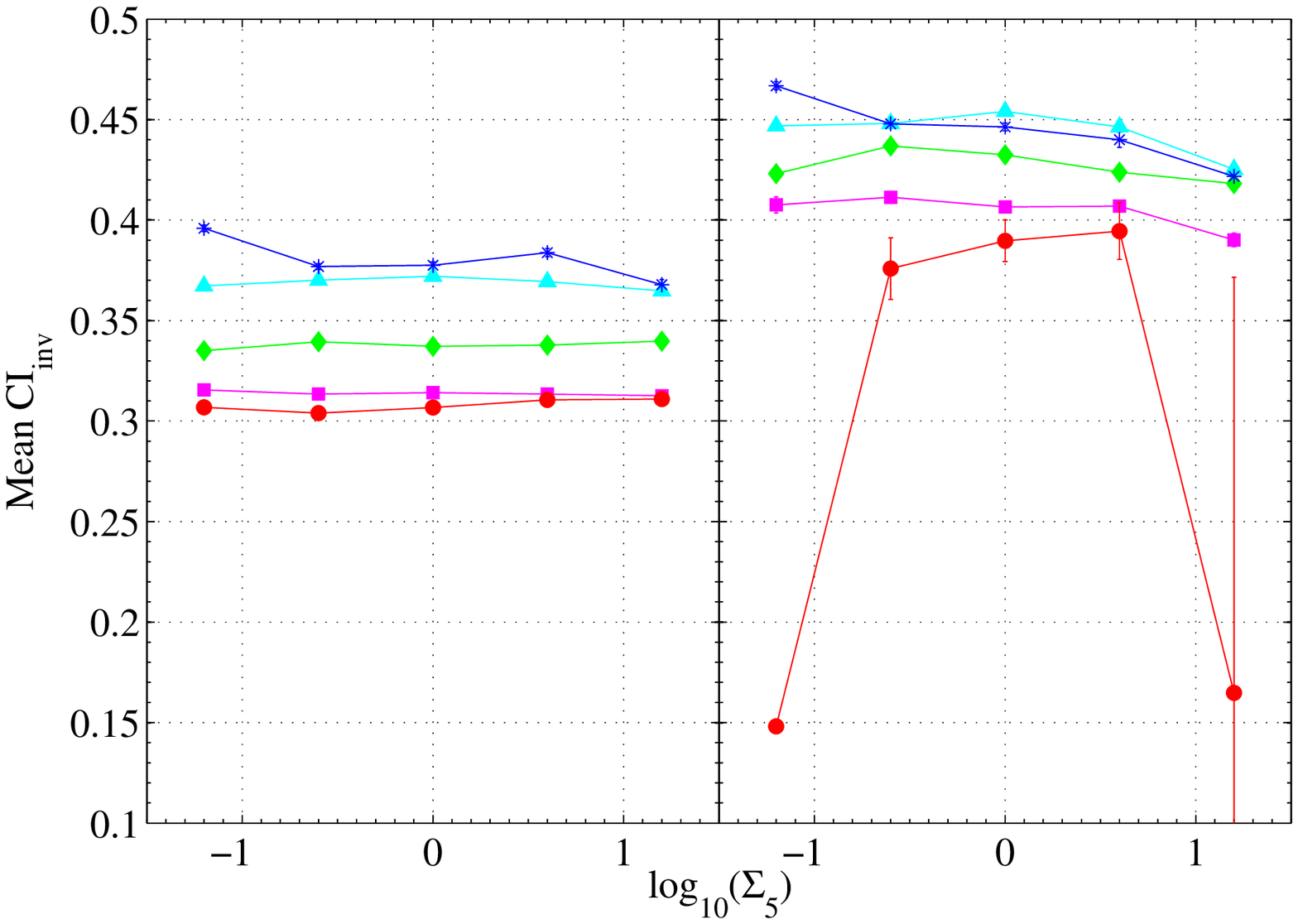}}}
\mbox{\subfigure{\includegraphics[width=3in]{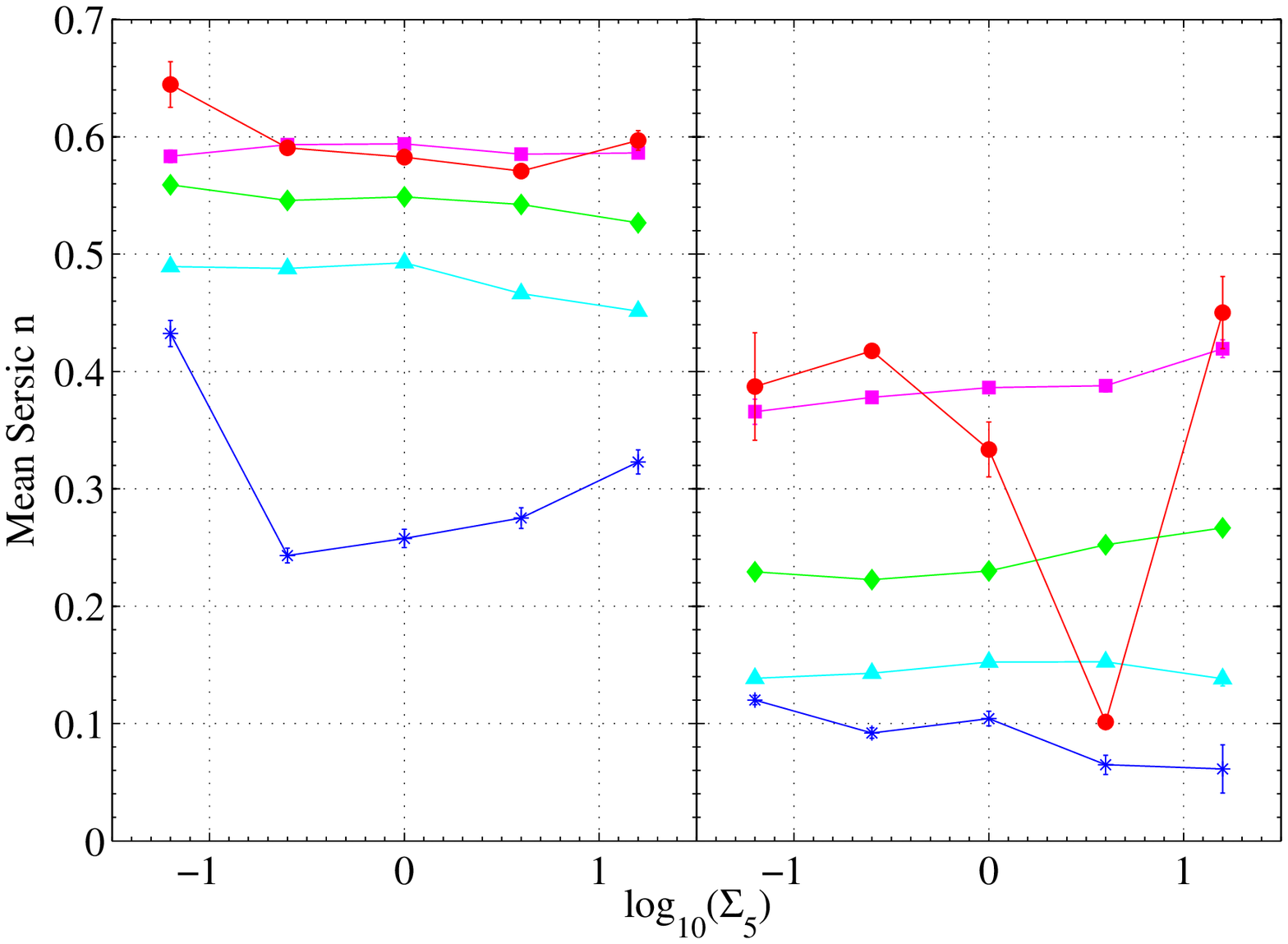}}\qquad
      \subfigure{\includegraphics[width=3in]{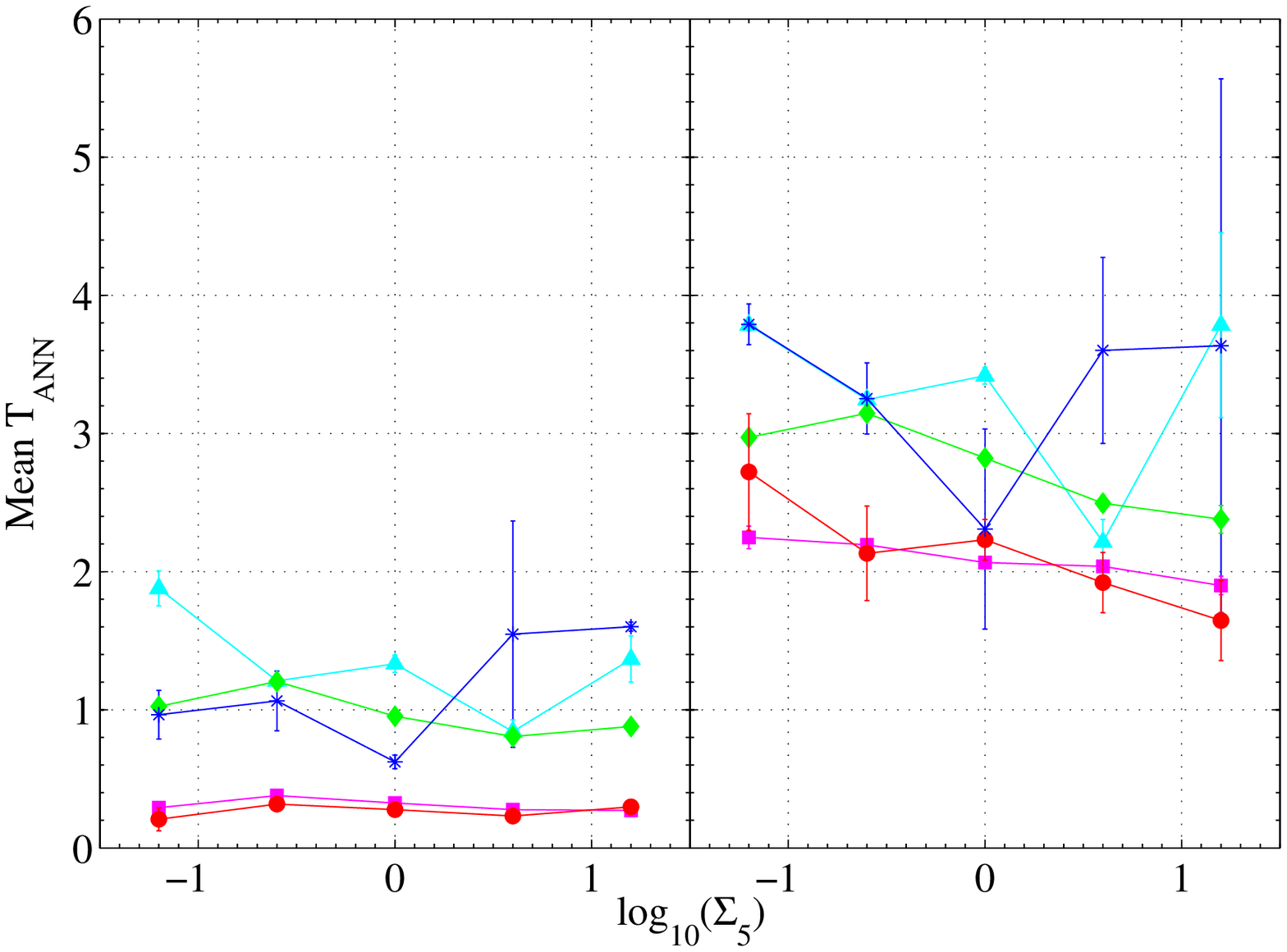}}}
\caption{Mean colour and type in each $M_r$ bin versus $\Sigma_5$ for $u-r$,
  $\ciinv$, S\'ersic $n$ and $\tann$. The left hand sides of each panel show
  red/early-type galaxies and the right hand sides show blue/late-type. The
  error bars are $1\sigma$. The left hand sides of the $u-r$ panel is similar to
  fig. 3 of B04. The magnitude bins are the same as for Fig. \ref{fig: Red
    fractions}. \label{fig: Mean colours}}
\end{figure*}

\begin{figure*} \centering
\mbox{\subfigure{\includegraphics[width=3in]{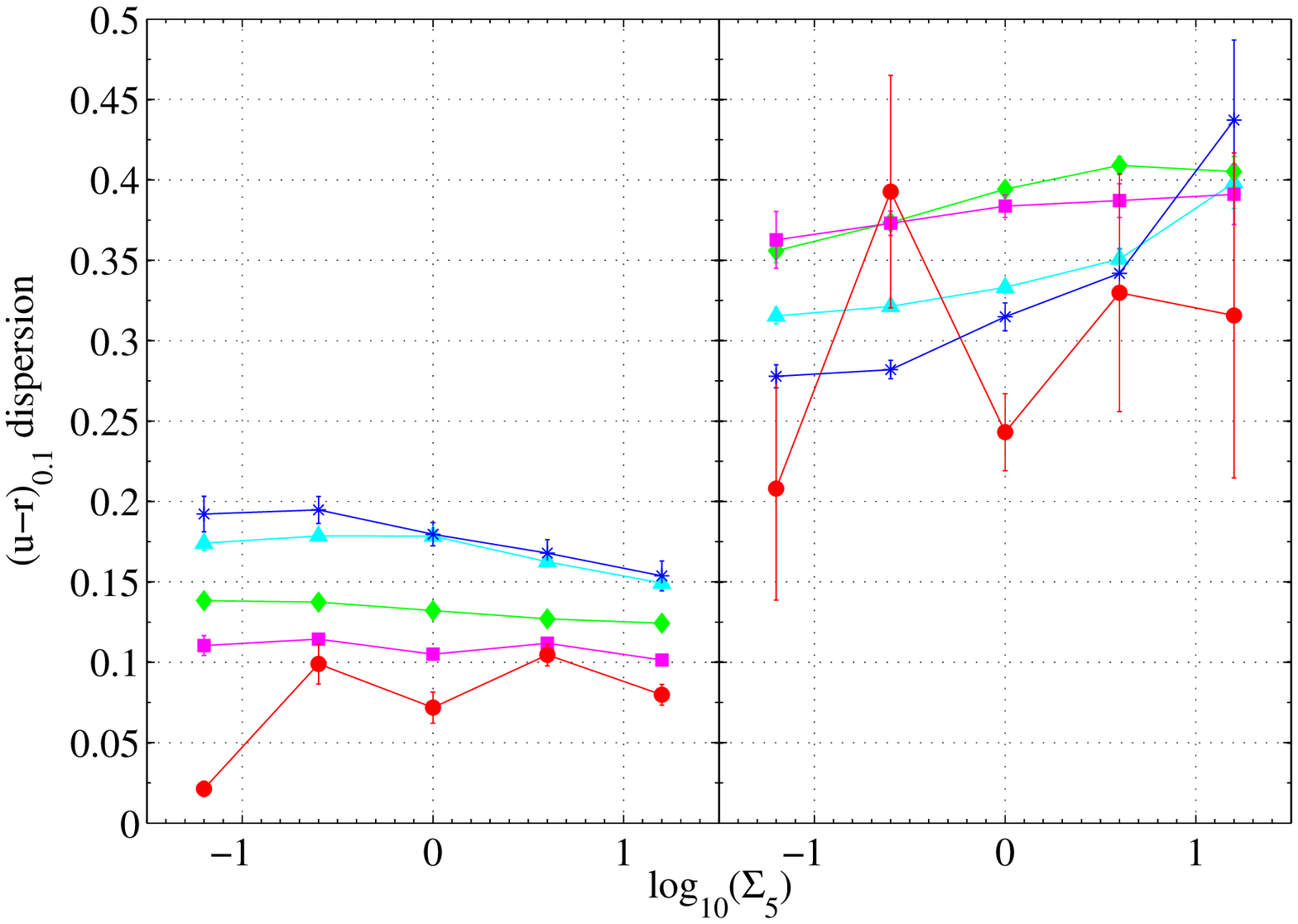}}\qquad
      \subfigure{\includegraphics[width=3in]{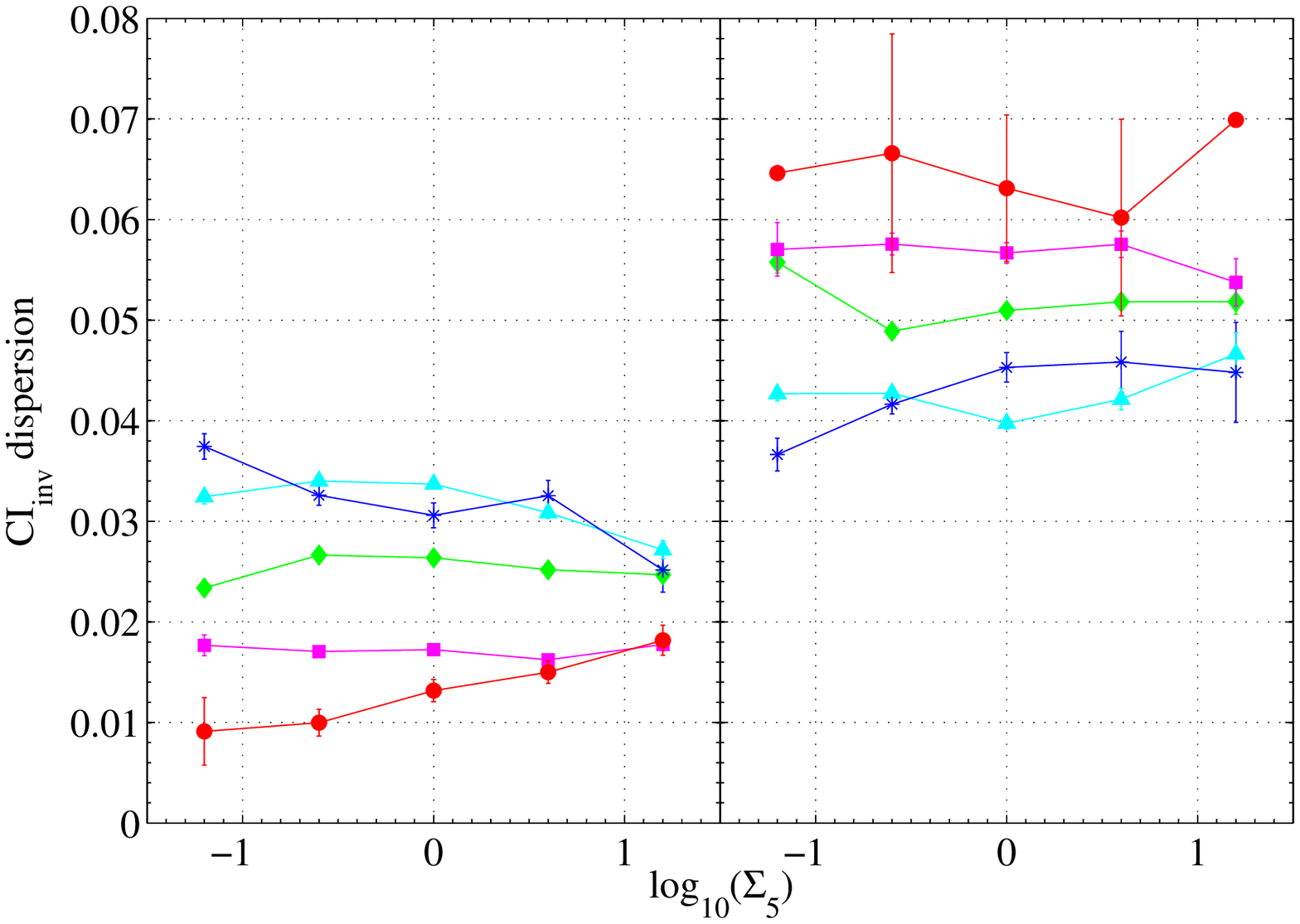}}}
\mbox{\subfigure{\includegraphics[width=3in]{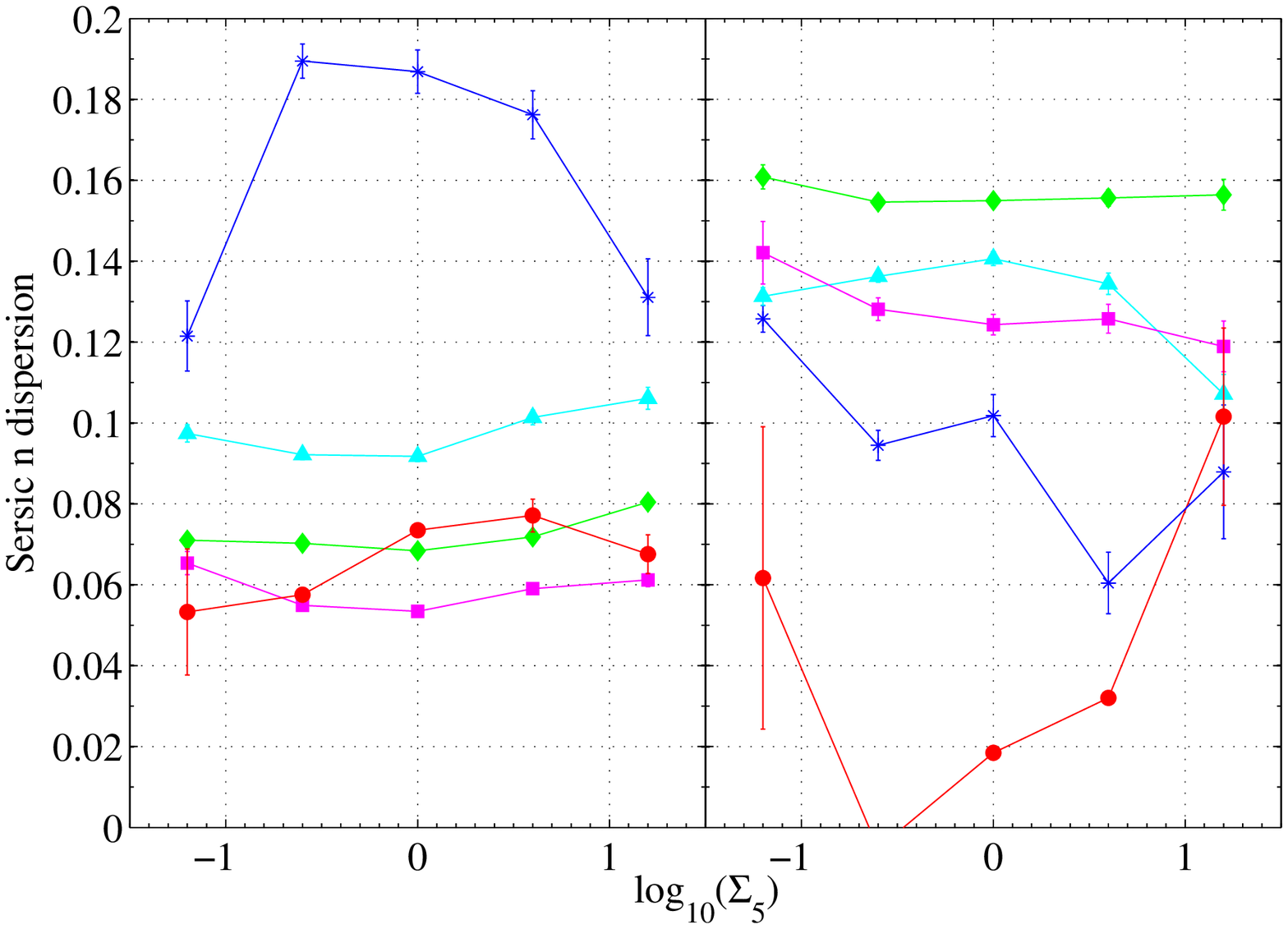}}\qquad
      \subfigure{\includegraphics[width=3in]{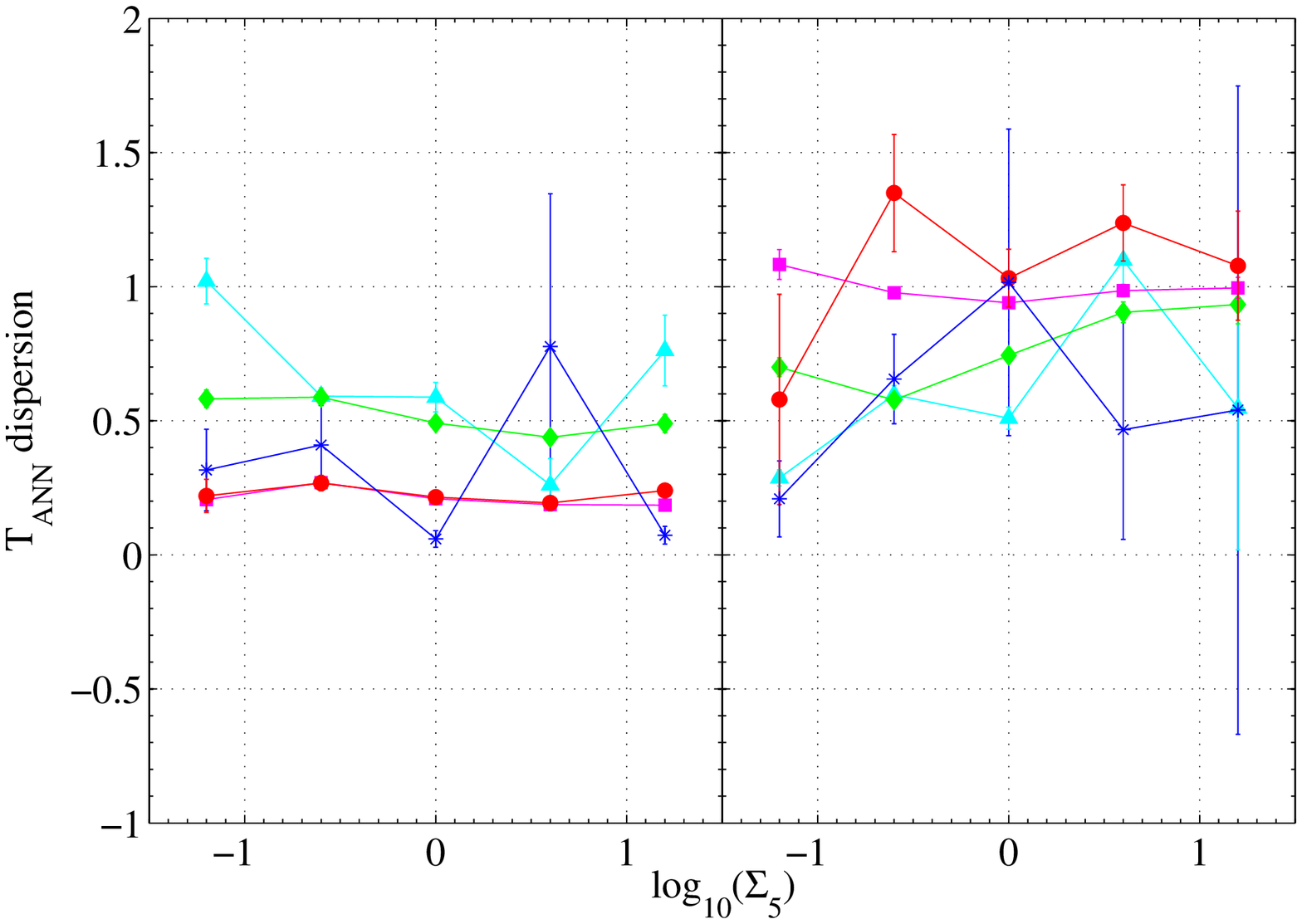}}}
\caption{As Fig. \ref{fig: Mean colours} but for the $\sigma$ dispersions in the
  Gaussian fits. \label{fig: Dispersions}}
\end{figure*}

Figs. \ref{fig: Red fractions}--\ref{fig: Dispersions} show some of the trends
in Figs. \ref{fig: CDL}--\ref{fig: MDLt} more clearly. The $\pm 1\sigma$ error
bars are calculated by finding the point at which the $\chi^2$ for the fit
increases by 1 around the minimum for each parameter in turn. The shape of the
$\chi^2$ is assumed to be parabolic, which is a good approximation, although it
assumes that the $\chi^2$ is minimized for the other parameters whilst the one
under study is being varied. However, these approximations will tend to
overestimate the errors, so it is good as a conservative estimate.

The top left panel of Fig. \ref{fig: Red fractions} shows the fraction of red
galaxies for $M_r$ versus $\Sigma_5$ for $u-r$. The fraction is given by the
ratio of the areas under the two Gaussians, which is given by $A/(A+B)$ assuming
that approximately the whole area under each Gaussian is populated by galaxies
(the normalised areas are 1). This is seen to be true in Figs. \ref{fig:
  CDL}--\ref{fig: MDLt}. The fraction rises with density, as expected, and
rises in a similar way over all luminosities, although the brightest galaxies at
$-22.5 < \mrlogh < -21.5$ are slightly noisy. The trends are similar to those in
the left hand panel of fig. 2 in B04.

The top left two panels of Fig. \ref{fig: Mean colours} show the mean $u-r$
colours $\overline x_1$ and $\overline x_2$ in a similar way for the two
Gaussians. The mean colour for the red galaxies is approximately independent of
density (as seen in Fig. \ref{fig: CDL}), with just a slight trend for redder
colours at higher densities. The blue galaxies show a more distinct reddening
towards the highest densities. This is similar to fig. 3 of B04.

The top left hand two panels of Fig. \ref{fig: Dispersions} show the $u-r$
dispersions $\sigma_1$ and $\sigma_2$. For red galaxies the dispersions are low
and for lower luminosities decline at the highest density, whilst those for the
blue galaxies are higher, slightly increasing at the highest density for low
luminosity. The dispersions for red galaxies also show a clear decrease with
increasing luminosity.

The remaining panels of Figs. \ref{fig: Red fractions}--\ref{fig: Dispersions}
show the same as $u-r$ but for the $\ciinv$, S\'ersic $n$ and $\tann$
morphologies. Each measure of morphology shows a similar behaviour. In
Fig. \ref{fig: Red fractions}, the early-type fraction shows some increase with
density in a similar way to the red fraction, whereas the late-type fractions
(not shown) are noisy.

In Fig. \ref{fig: Mean colours} the mean type for the early-type galaxies is
seen to be approximately independent of density at fixed luminosity in the same
way that colour was for the red galaxies. This hints that the same population is
being seen, which is consistent with a distinct bright, concentrated, early
spectral type, red population. The late types change little with density, the
$\ciinv$ and $\tann$ becoming earlier but the S\'ersic $n$ showing no trend. The
similarity of the patterns seen suggests that in terms of the effect of
environmental density, colour is closely correlated with morphology, but that
there could be a residual relation.

Fig. \ref{fig: Dispersions} shows that for morphology the dispersions do not
show any obvious trends with density. For $u-r$, this justifies the constraint
in B04 for the the dispersion not to vary. (B04 also find little change if they
relax this constraint.) Some of the larger changes are due to the widths of
individual Gaussians being poorly constrained by either significant overlap
between them or by low numbers of galaxies.

\subsection{Residual Colour- and Morphology-Density} \label{subsec: residual} 

The residual relations are calculated by binning the galaxies into a
2-dimensional grid of colour and morphology, using the same histogram bins as
in Figs. \ref{fig: CDL}--\ref{fig: MDLt}. The mean density $\mu_{ij}$ in each bin
is calculated, along with the marginal averages as a function of colour bin
($\mu^c_i$) and of morphology bin ($\mu^m_j$), using the densities for the
individual galaxies in each so that the bins are correctly weighted. In order to
factor out the colour-dependence of density, one divides each mean density
$\mu_{ij}$ by the appropriate marginal average $\mu^c_i$. A re-calculation of
the marginal dependence of density on colour would then give all ones, whereas a
re-calculation of the marginal dependence on morphology will yield the residual
morphology-dependence of density after that due to colour has been removed. By
instead dividing out the morphology-dependence in an equivalent way, the
residual colour dependence may be calculated. The errors on each point are the
$1\sigma$ dispersion of the values of the bins which are averaged to give that
point.

Fig. \ref{fig: resid colour} shows the residual relation for $u-r$ on removal
of morphology. There is a definite trend in density with colour, consistent with
numerous earlier studies \citep[e.g.][]{kauffmann:envt,blanton:envtbbprops} in
which the colour was a strong predictor of environmental density. As expected,
the density increases for redder galaxies. The error bars are still relatively
large, reflecting the less certain nature of $\Sigma_5$ compared to colour.

Fig. \ref{fig: resid morph} shows the morphology-density relation after the
removal of the change in density due to the change in colour with morphology. In
contrast to colour, the residual relation is consistent with a value of one,
i.e. no residual relation. This means that there is no evidence for a change in
morphology that is not correlated with a change in colour.

\begin{figure} \centering
\includegraphics[width=3in]{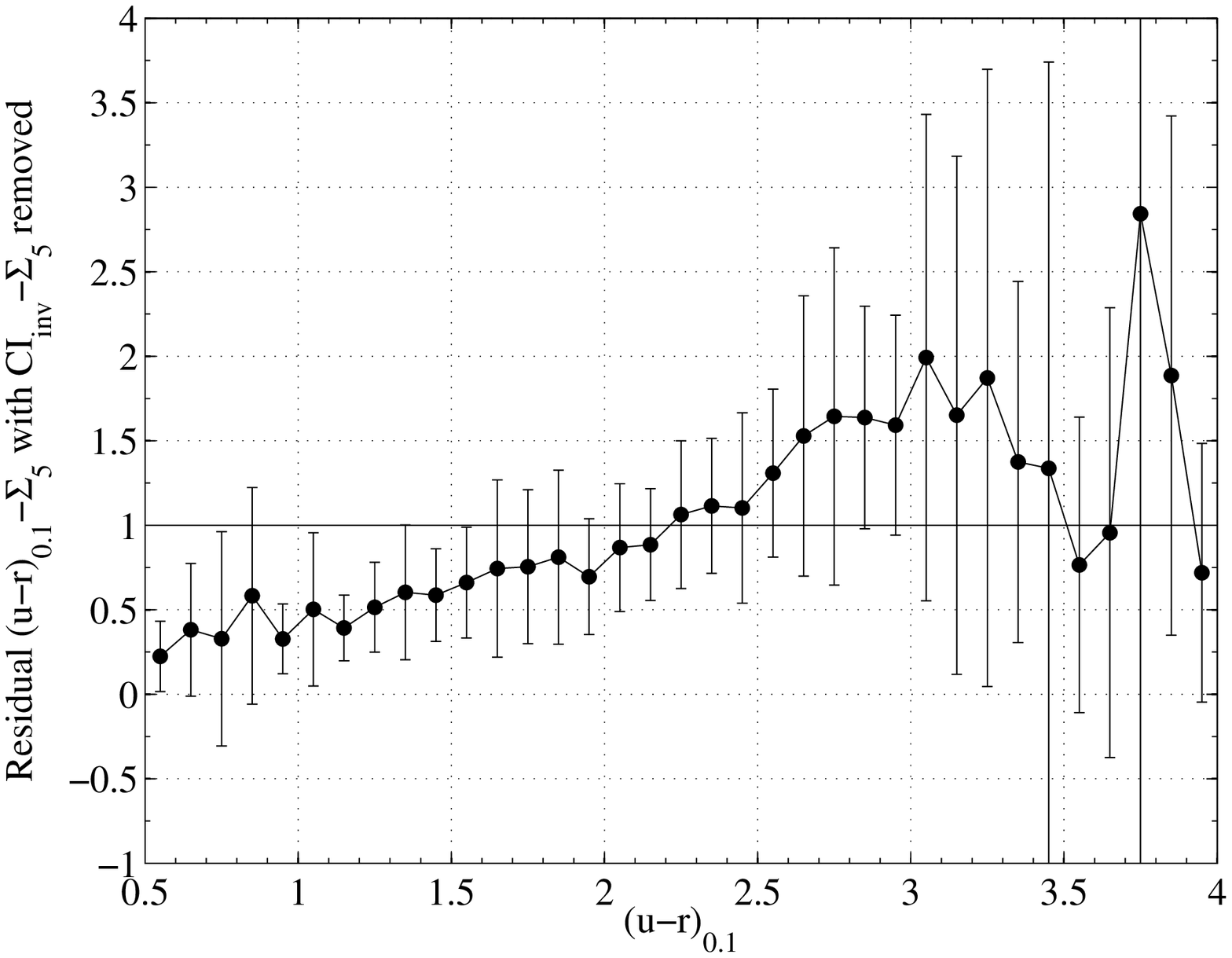}
\includegraphics[width=3in]{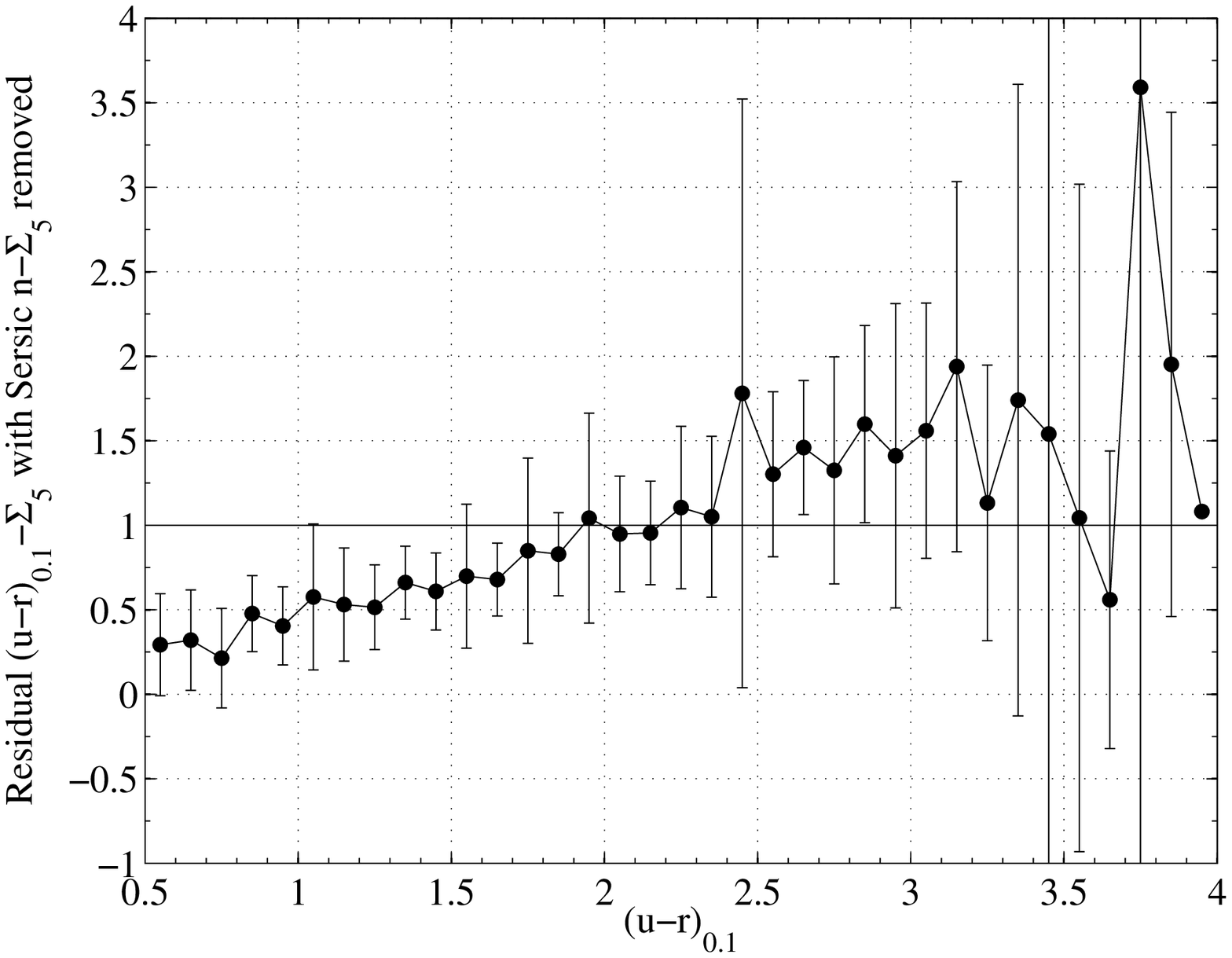}
\includegraphics[width=3in]{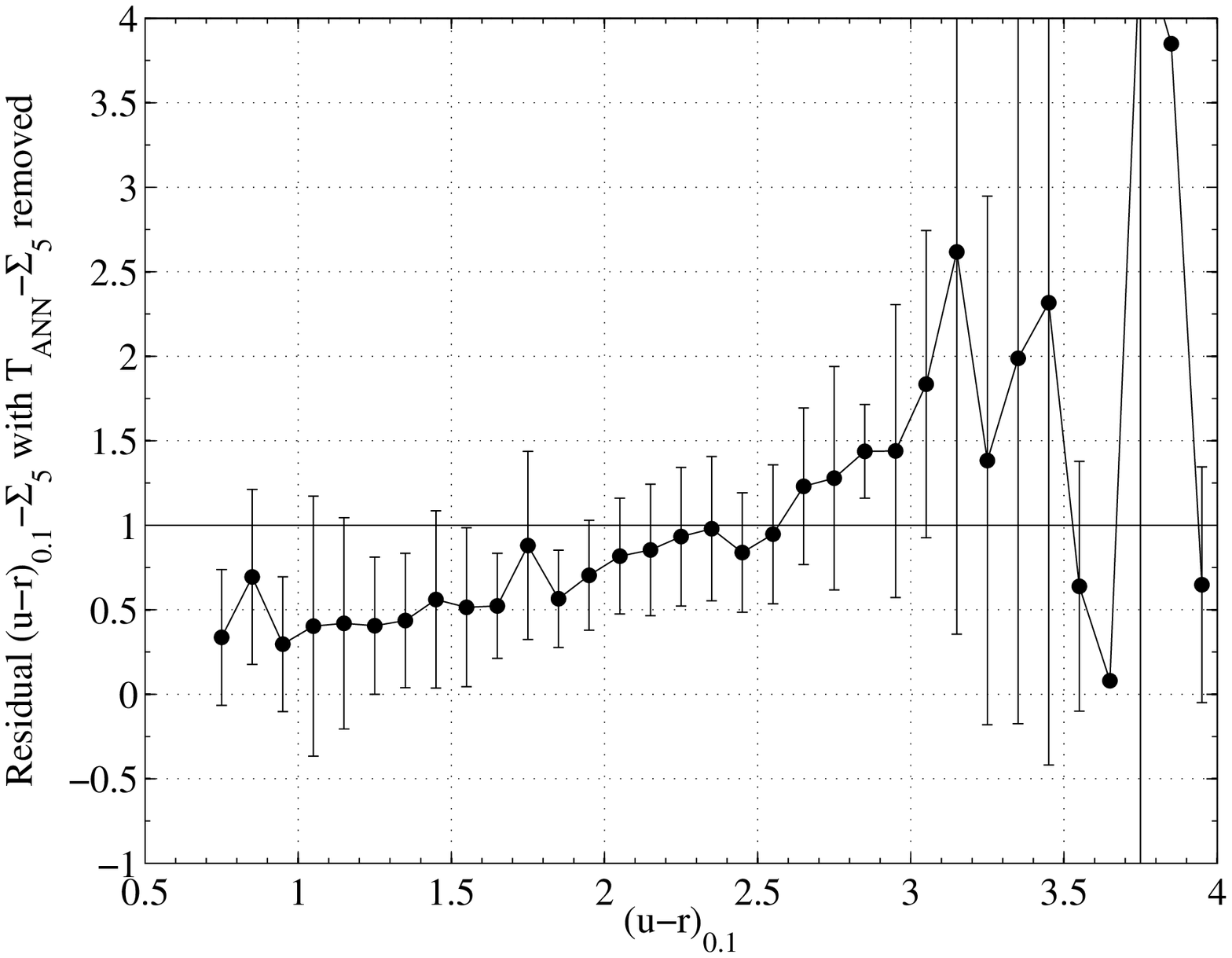}
\caption{Residual density dependence of colour once the dependence due to
  morphology ($\ciinv$, S\'ersic $n$ and $\tann$ in respective panels) has been
  removed. The sample sizes are 79,495, 75,753 and 13,638 galaxies
  respectively. The error bars are the $1\sigma$ dispersion in the morphology
  bins, and a value of 1, highlighted by the horizontal line, indicates
  no residual. There is a clear reddening of colour with density. \label{fig:
    resid colour}}
\end{figure}

\begin{figure} \centering
\includegraphics[width=3in]{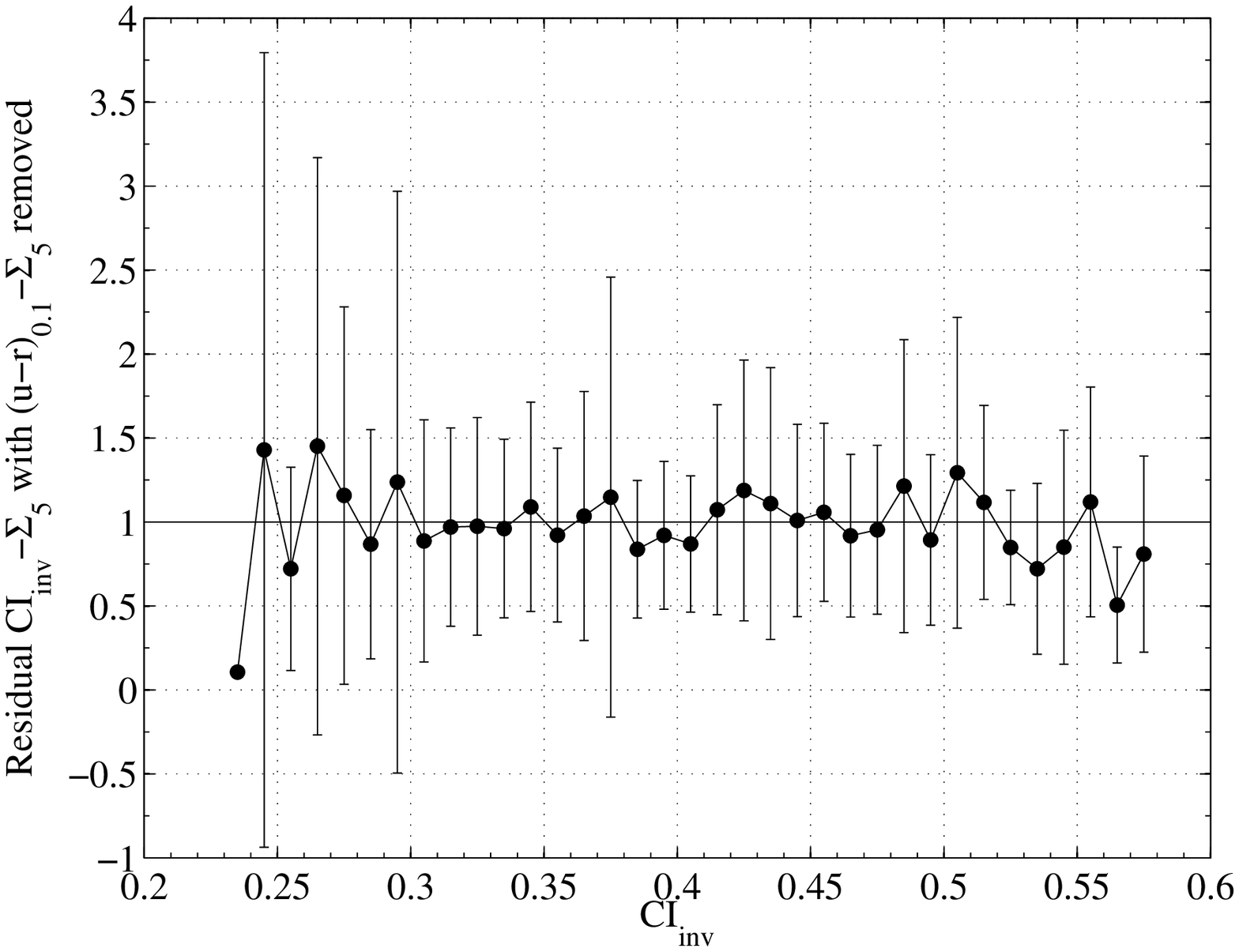}
\includegraphics[width=3in]{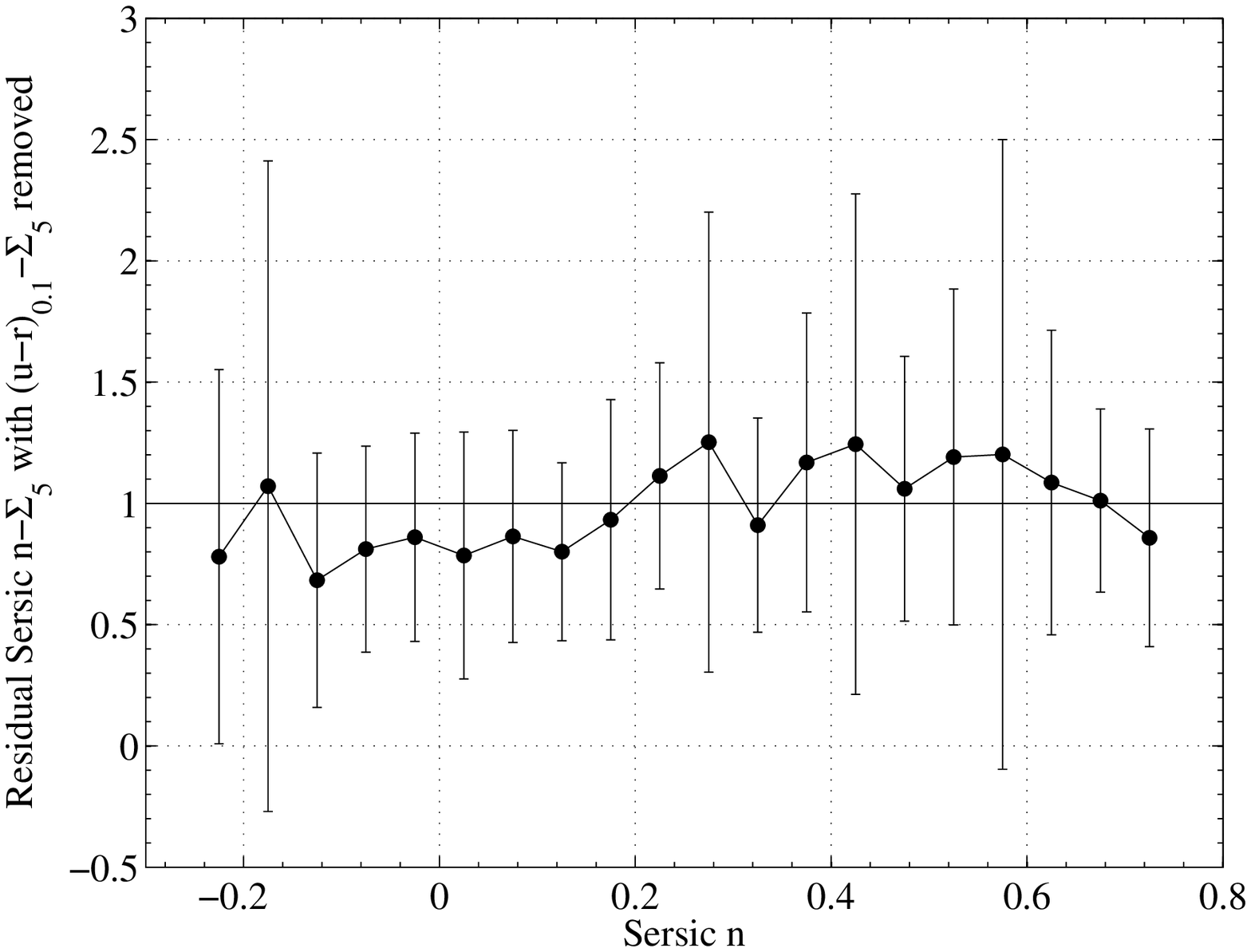}
\includegraphics[width=3in]{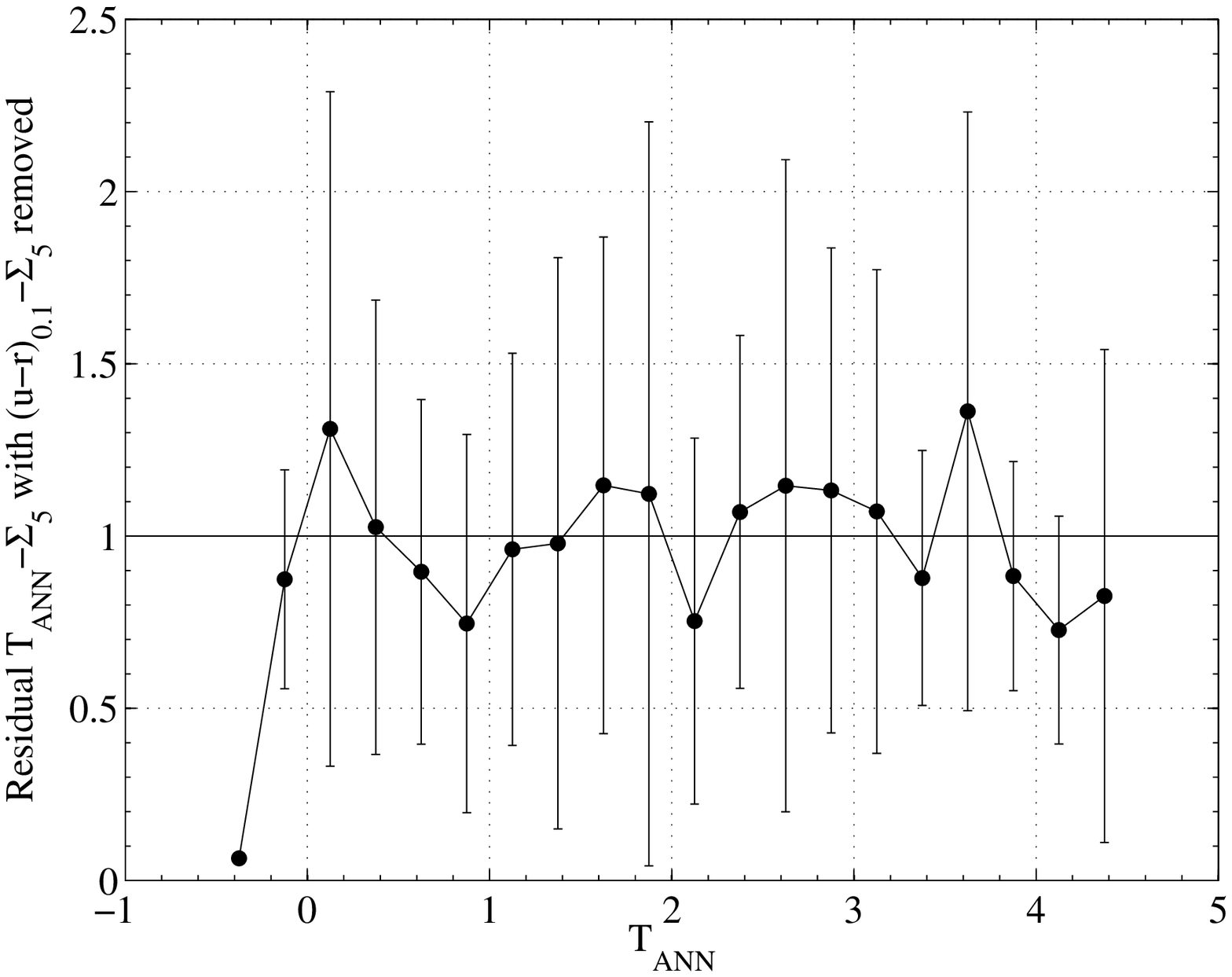}
\caption{As Fig. \ref{fig: resid colour}, with the same sample sizes, but for
  the residual density dependence of morphology once that due to colour has been
  removed. Unlike colour, here there is no clear trend. \label{fig: resid morph}}
\end{figure}

\subsection{Comparison to Previous Work} \label{subsec: comparison} 

Besides B04, who presented the colour-density relation for the EDR and for whose
results ours broadly represent an update to the SDSS DR4 (\S \ref{subsec:
  fraction} and \S \ref{subsec: residual} above), substantial work involving
similar measures of colour and morphology with comparison to density has been
carried out by \citet{goto:md} and \citet{park:envt}.

\citet{goto:md} present the morphology-density relation for the SDSS EDR. They
assign morphology using the inverse concentration index, as used here, and a
preliminary version of the texture parameter described by
\citet{yamauchi:citexture}. The latter is corrected for elliptical isophotes and
gives an improved correlation to visual morphology. The galaxies are divided
into early, intermediate, early disc and late disc. The sample is volume-limited
to $\mrlogh < -19.9$ in the redshift range $0.05 < z < 1$ and contains 7,938
galaxies.  This compares to our values of $\mrlogh < -19.5$, $0.001 < z <
0.0889$ and 13,655 galaxies. The same $\omegam = 0.3$, $\omegal = 0.7$ Euclidean
cosmology is used. Their density estimator is three dimensional.

They find the expected trend of increasing early-type fraction with increasing
density, and also two characteristic scales, giving three density regimes. In
the lowest density regime, the relation is less noticeable. In the intermediate
density regime, intermediate type galaxies increase with density and spirals
decrease. In the densest regime, the intermediate types decrease and the early
types increase. Thus the morphology is little affected in the low density
regime, then subject to a mechanism which inhibits star formation and turns
spirals into intermediate types in the intermediate density regime, then a third
process decreases these types and increases early types in the dense regime. Our
work is consistent with this, with the bridge between the two bimodal
populations resembling the intermediate types described here. It is difficult to
compare quantitatively, however, as the density measures are different.

\citet{park:envt} study numerous galaxy properties as a function of environment
in various volume-limited samples containing up to 80,479 galaxies for $0.025 <
z < 0.11$ using a subset of the SDSS DR5. The galaxy properties studied include
the $(u-r)_{0.1}$ colour and morphology assigned using the colour gradient
method of \citet{park:segregation}. Their measure of density is obtained using
an adaptive smoothing kernel, so again an exact quantitative comparison with our
results would be difficult. Their fig. 9 shows the fraction of early-type
galaxies (as defined in \citet{park:segregation}) as a function of density and
luminosity. The result shows the same trends as our Fig. 5, with a monotonic
increase in early-type fraction with density and luminosity, and the trend lines
separated by luminosity approximately parallel to each other. However, they also
fix both luminosity and morphology and then find little residual dependence of
properties on environment, except for a sharp decrease in late types at the
bright end.

\section{Discussion} \label{sec: MD Discussion} 

Considering the number of galaxies, the variety of their properties and
environments and the quality of the SDSS data, the rest-frame colour histograms
are well fit by a simple sum of two Gaussians. This result is also seen by B04
and others and strongly suggests two underlying populations of galaxies. This is
consistent with the patterns seen in the LFs in B06 (their figs. 2, 3, 4 and 6
show  the LF bivariate with $\tann$, $\ciinv$, S\'ersic $n$ and $(u-r)_{0.1}$
respectively). There, an early morphological type, concentrated, early spectral
type, red, high S\'ersic index population of galaxies is evident and the LFs
binned in absolute magnitude and plotted against their second parameter often
show bimodality.

One of the main questions asked in this paper is whether the morphology reveals
anything that the colours do not. Compared to $u-r$, the $\ciinv$ and S\'ersic
$n$ also show clearly bimodal distributions, although the $\ciinv$ plot becomes
unimodal at fainter magnitudes and low densities. There are no obvious residual
relations when those due to colour are removed, thus these measures of
morphology appear not to be predictive of environment beyond their correlation
to colour.

When the $\tann$ morphology-density relation is plotted the distributions are
less bimodal, ranging from looking similar to the plots for colour to being not
bimodal at all. There are various possible reasons for this, including (1) the
morphology may be less sensitive to the processes producing the bimodal
population in colour, being intrinsically spread when compared to the
colour-producing physics; (2) there may be more than two populations, the third
or higher to which colour may not be sensitive (an example might be S0 galaxies,
which are almost as red as ellipticals); or (3) the neural network morphological
types might be spread or biased versus the true types. The last point is not
thought to be likely \citep{ball:ann}. It is difficult to distinguish between
the intrinsic spread or more populations possibilities, although the fact that
some of the bins are more bimodal than others favours the third population
idea. Here neither possibility is ruled out and indeed both could be
occurring. More detailed investigation of the morphology-colour-density
relationship may be useful, but it may remain the case that the morphologies are
simply intrinsically too `fuzzy' to show the underlying populations as clearly
as the colour. That said, $\tann$ has larger error bars, so the formal goodness
of fit is  not necessarily worse than that for $u-r$, $\ciinv$ and $n$.

This spreading of Hubble types between the two populations is also seen by
various other workers. In the Millennium Galaxy Catalogue
\citep[MGC,][]{liske:mgclf}, \citet{driver:mgcmorph} study a sample of 3314
galaxies with eyeball Hubble types at $B < 19$ and find that, while types E/S0
(bulges) exclusively occupy the early-type peak and Sd and later (discs) the
late-type peak, the Sa--Sc, which are a mixture of bulge and disc, are spread
between the two peaks. Also in the MGC, \citet{ellis:subdivision} study a
well-defined sample of 350 galaxies and find, using multivariate statistical
analysis of the galaxy properties, that only two populations, corresponding to
early and late types, are justified, and the late types (Sa--Irr) are smoothly
spread within the late-type region of parameter space.  In a sample of 1,246
galaxies with Hubble types from the Revised Shapely-Ames catalogue,
\citet{vandenbergh:envt} similarly find a continuum of Hubble types but a
bimodality in colour.

In a sample of 22,121 galaxies from the Third Reference Catalogue of Bright
Galaxies, \citet{conselice:classification} finds that the best three parameters
with which to classify galaxies are mass, star formation and merger history. The
Hubble type correlates most strongly with stellar mass, $(U-B)_0$ and $(B-V)_0$
colour. They also (his fig. 10) confirm the intrinsic spread in morphology
when compared to other properties such as colour, showing that morphology is at
least to some extent independent of the other parameters.

\citet{baldry:bimodalrev} discuss the work in \citet{baldry:bimodalcmd} and B04
and agree with the interpretation of morphology, in the sense that `whatever
processes give rise to the blue/red distribution should also give rise to
\emph{distributions} in morphology', although they go on to say that `thus, S0
or Sa galaxies could have a probability of belonging to one or the other
distributions and should not be considered as classes'.

The physics giving the bimodal population in colour is thought to be that red
galaxies have a passively evolving old stellar population and are formed by
major mergers. The blue galaxies are undergoing star formation and less violent
accretion events. Galaxies can change from blue to red and when they do they do
so rapidly, leaving few galaxies in between the two populations. B04 go on to
propose that `most star forming galaxies today evolve at a rate that is
determined primarily by their intrinsic properties, and independent of
environment', which is consistent with the results here. The morphology may be 
seeing extra physics at work, for example ram-pressure stripping of gas from a
galaxy infalling into a cluster can turn a spiral into an S0 rather than an
elliptical, even though they are almost as red. S0s are sufficiently common that
they should show up as a component in the plots in this paper if they indeed
form a component distinct from other galaxies. In comparison to simulations,
only recently have semi-analytic models begun to produce bimodal populations
matching those observed \citep[e.g.][]{cattaneo:bimodality}, so there may still
be much to be gained by a detailed intercomparison between the two.

The ANN types are biased away from very early or late types, but not to an
extent large enough to affect the overall trends seen here. This is thought to
be due to the importance of the concentration index in the training set, which
has a similar bias. A larger training set for morphology is needed, particularly
for very late types of type Scd or later, corresponding to $\tann \geq 4.5$. Our 
training set was based on the SDSS EDR version of the catalogue of
\citet{fukugita:morph}. However, their published version, while based on DR3 and
40\% larger at 2658 galaxies instead of 1875, would be unlikely to significantly
improve our results due to the intrinsic spread in $\tann$ and the $r$-band
object selection of the SDSS, which limits the Scd or later types to 5.2\% of
the sample. They compare this to 10.6\% from the earlier B-band study of
\citet{fukugita:baryon}.

Here the model $u-r$ colour was used. \citet{driver:mgcmorph} showed that the
core $u-r$, measured using the PSF colour, provides a cleaner separation for
their sample when used in conjunction with the S\'ersic index. The same may be
true here if that $u-r$ were used, although the model $u-r$ is cleaner than the
Petrosian $u-r$, for which we also generated results. \citet{driver:mgcmorph}
also argue that the two fundamental components are, rather than red and blue
galaxies, bulges and discs. This is consistent with the results here. However,
it is not necessarily that simple, as \citet{drory:pseudobulge} differentiate
classical and pseudobulges, which have different formation mechanisms, and find
that, for types Sa-Sbc for a given bulge-to-total light ratio, the former are
predominantly in red galaxies and the latter in blue, i.e., the bimodal
distribution is not simply due to the bulge to disk ratio.

\section{Conclusions} \label{sec: MD Conclusions} 

Galaxy properties are studied as a function of environment in the
SDSS, specifically the relations between $u-r$ restframe colour, morphological
type and the environmental density. The colour is given by the restframe value
of $u-r$. The morphology is in the form of the inverse concentration index,
$\ciinv$, the S\'ersic index, $n$ and the Hubble T type, $\tann$, the latter 
assigned by a trained artificial neural network. The density is the $\Sigma_5$
surface density of galaxies measured by the fifth nearest neighbour brighter
than an $r$ band absolute magnitude of $\mrlogh = -19.5$. The range of density
probed covers all environments from the cores of rich clusters to the field. The
colour results are similar to those of \citet{balogh:bimodallfenvt} (B04) 
but here these are extended to galaxy morphology.

Subdivision of the population by density and luminosity shows a clear bimodal
distribution in the colour-morphology plane. For each subsample we fit the colour and
morphology by a sum of two Gaussians. The well-known colour-density and
morphology-density relations are seen. The colour is well fit by the Gaussian
sum, confirming two underlying populations, one red and one blue, as seen by
B04. The mean colour of the red population is approximately constant but the
blue population gets slightly redder with increasing density. The relations are
seen at high significance over the range of densities probed, thus confirming
that they extend into the field and are not just present in galaxy clusters. 

Compared to $u-r$, $\ciinv$ and $n$, the $\tann$ morphology is less obviously
fit by the sum of two Gaussians, suggesting one or more of the possibilities (1) 
three or more populations (e.g. early, lenticular and late), (2) sensitivity of
the morphology to processes which do not affect the colours, or (3) a higher
intrinsic spread in morphology compared to two underlying populations and
physical processes.

While the $\ciinv$ and S\'ersic $n$ to some extent separate the two main galaxy
populations, certainly more cleanly than the Hubble type, both still suffer from
extensive overlap in the populations. Figs. \ref{fig: CI contour}--\ref{fig: T
  contour} show that, rather than using a single colour or morphology, improved
separations should be obtained using a division in the plane of colour and
morphology rather than any single measure. In its simplest form in
Figs. \ref{fig: CI contour}--\ref{fig: T contour}, this would correspond to the
allowance of an oblique line as a separator instead of an axis-parallel line
(c.f. e.g. fig. 6 of \citet{choi:properties}, fig. 10 of
\citet{baldry:massenvt}, or fig. 15 of \citet{driver:mgcmorph}, which also show
this planar distribution).

The colour-density relation divided into bins by absolute magnitude reproduces
the results of B04: the mean colours are independent of density for fixed
luminosity but the red fraction increases with increasing density. The fraction
of red galaxies increases approximately linearly with density, with parallel
slopes of lower fraction for decreasing luminosity. The dispersion in colour for
each population is low for red galaxies, increasing with decreasing luminosity
and decreasing with density for low luminosity. For blue galaxies it is
significantly higher increasing at low luminosity with density. At high
luminosity there is now a significant trend with density.

The morphology-density relations divided in this way are similar if the early-type
galaxies are red. The early-type galaxy fractions increase in a similar way to
the fraction of red galaxies but the trends are noisier. The mean types versus
density are similar to colour but the late types do not as obviously get earlier
with density, consistent with the colour having a stronger dependence on density
than morphology. The dispersions show few significant trends with density.

In $u-r$ there is a clear residual relation of galaxies for both blue and red
galaxies with density even after the reddening due to the increase in early
types is removed. This suggests that colour has a stronger dependence on density
than morphology, in agreement with other studies.

For each of $\ciinv$, $n$ and $\tann$, there is no evidence of a residual trend
once that due to variation with colour is removed. Therefore, for these data,
none of the measures of morphology are predictive of environment beyond their
correlation to colour.

\section*{Acknowledgments}

Nick Ball thanks Bob Nichol, Ivan Baldry, and Michael Balogh for useful
discussions, and K. Simon Krughoff and Chris Miller for help with the DR4 VAC
catalogue. We thank the referee for a comprehensive and constructive report
which improved the paper.

\noindent NMB was funded by a PPARC studentship from 2001--2004 and, with RJB, 
would like to acknowledge support from NASA through grants NN6066H156 and
05-GALEX05-0036, from Microsoft Research, and from the University of
Illinois.

Funding for the SDSS and SDSS-II has been provided by the Alfred P. Sloan
Foundation, the Participating Institutions, the National Science Foundation, the
U.S. Department of Energy, the National Aeronautics and Space Administration,
the Japanese Monbukagakusho, the Max Planck Society, and the Higher Education
Funding Council for England. The SDSS Web Site is http://www.sdss.org/.

The SDSS is managed by the Astrophysical Research Consortium for the
Participating Institutions. The Participating Institutions are the American
Museum of Natural History, Astrophysical Institute Potsdam, University of Basel,
Cambridge University, Case Western Reserve University, University of Chicago,
Drexel University, Fermilab, the Institute for Advanced Study, the Japan
Participation Group, Johns Hopkins University, the Joint Institute for Nuclear
Astrophysics, the Kavli Institute for Particle Astrophysics and Cosmology, the
Korean Scientist Group, the Chinese Academy of Sciences (LAMOST), Los Alamos
National Laboratory, the Max-Planck-Institute for Astronomy (MPA), the
Max-Planck-Institute for Astrophysics (MPIA), New Mexico State University, Ohio
State University, University of Pittsburgh, University of Portsmouth, Princeton
University, the United States Naval Observatory, and the University of
Washington. 

This research has made use of NASA's Astrophysics Data System.




\label{lastpage}

\end{document}